\documentclass{article}

\usepackage{arxiv}

\usepackage[utf8]{inputenc}
\usepackage{tabularx}
\usepackage[figuresright]{rotating}
\usepackage{amsmath,amssymb,bbm}
\usepackage{relsize}
\usepackage{graphicx}
\usepackage{arydshln}
\usepackage{booktabs}
\usepackage{multirow}
\usepackage{natbib}
\usepackage{float}
\usepackage{appendix}
\usepackage{algorithm}
\usepackage{makecell}
\usepackage{threeparttable}
\usepackage{setspace}

\usepackage[noend]{algpseudocode}
\DeclareMathOperator{\Tr}{Tr}
\newcommand{\te}[1]{\text{#1}}
\newcommand{\ti}[1]{\textit{#1}}

\newcommand{\E}[2]{\mathbbm{E}_{#1}\left[{#2}\right]}

\newcommand\numberthis{\addtocounter{equation}{1}\tag{\theequation}}

\title{Sparse Dynamic Factor Models with Loading Selection by Variational Inference}

\author{
  Erik Sp{\aa}nberg \\
  Department of Statistics\\
  Stockholm University\\
  SE-106 91, Stockholm \\
}
\bibpunct[, ]{(}{)}{;}{a}{,}{,}

\begin{document}
\maketitle

\onehalfspacing
\begin{abstract} 
    In this paper we develop a novel approach for estimating large and sparse dynamic factor models using variational inference, also allowing for missing data. Inspired by Bayesian variable selection, we apply slab-and-spike priors onto the factor loadings to deal with sparsity. An  algorithm is developed to find locally optimal mean field approximations of posterior distributions, which can be obtained computationally fast, making it suitable for nowcasting and frequently updated analyses in practice.  We evaluate the method in two simulation experiments, which show well identified sparsity patterns and precise loading and factor estimation. 
\end{abstract}


\section{Introduction}
In this paper, a variational inference (VI) algorithm is introduced to estimate dynamic factor models (DFMs) with loading selection, which can be used to find sparse solutions or to form structural specifications in large scale data analysis.

Imagine a large data set which is driven by a collection of unobserved common factors. It represents a description of reality where what we observe is projected by some deeper pattern. Typically, we imagine that this pattern belongs to a smaller dimension than the data itself, i.e the size of data cross-section exceeds the number of factors.  There is power to this idea;  if we are able to satisfactorily estimate such underlying factors, then we have obtained a manageable analyzing tool, with possible predictive capabilities. This is the approach taken by a DFM. 

The distinctive feature of a DFM, in relation to other factor models, is that it explicitly designs factors to be dynamic and have dynamic influence over the data. DFMs were introduced independently by \cite{SargentSims1977} and \cite{Geweke1977}, and over the last several decades extensive literature on the subject has developed, ranging from different estimation methods, theoretical qualities and practical applications. Some results worth of particular mention are consistent estimation in approximate models \citep{ForniEtAl2000, ForniEtAl2004, ForniEtAl2005}, identification theory \citep{BaiWang2014}, estimation by Expectation Maximization \citep{RubinThayer1982, WatsonEngle1983, BanburaetAl2014, Spanberg2021a} and Bayesian techniques \citep{OtrokWhiteman1998}. See review by \cite{BaiWang2016}. 

If we define a time series $\ti{y}_{i,t}$, with variable index $i$ at time $t$, a factor model can be described by
\begin{align*}
    \ti{y}_{i,t} = X_{i,t} + \epsilon_{i,t}, \quad t=1, ..., T, \quad i = 1, ..., n
\end{align*}
where $\epsilon_{i,t}$ is a idiosyncratic component and $X_{i,t} = \lambda_i'F_t$ is a common component, given by a vector of factor loadings, $\lambda_i$, and common factors $F_t$. In an \textit{exact} factor model, the idiosyncratic components are independent across $i$-index. Idiosyncratic components are often modelled as Gaussian distributed. 

More specifically, in DFMs, common factors are explicitly given a dynamic form, according to $F_t = [f_t' \; f_{t-1}' \; ... \; f_{t-p}']'$, where $f_t$ is a vector of dynamic factors at time $t$. They are popular in nowcasting \citep[see e.g.][]{BanburaEtAl2011,BanburaRunstler2011}, due to their ability to handle missing data and publication delays in large data sets. \cite{BanburaetAl2014} develops a maximum likelihood point-estimator using Expectation Maximization \citep[see][]{DempsterEtAl1977}, which can deal with any arbitrary missing data pattern. Building on their work, \cite{Spanberg2021b} estimates DFM using VI, dealing with missing data in a similar manner, but also considers parameter uncertainty and parameter shrinkage by posterior approximation. Their method shows promising approximation precision, which can be achieved very computationally fast relative to standard Markov Chain Monte Carlo (MCMC) procedures, making it practically viable for professional forecasters' daily work. 

The benefit of factor models in big data situations is apparent, however, such sizes come with considerations. When applying DFMs to hundreds or even thousands of variables, many of them might be redundant or even interfering the goal of finding some particular features in the data; some things are irrelevant and other things are highly relevant.  Ideally, we would like to perfectly sort out the factor loadings of interest. This is not always solely a question of empirical objectivity, but a question of what is the particular aim of the data analysis. We are perhaps more interested in some features of the data than others.

With this in mind, a procedure which can a-priori, but not dogmatically, direct the analysis to specific features could be highly valuable in analytic and predictive pursuits. Regardless if we are agnostic or dogmatic about which features are of interest, we still want to sort out the relevant factors and variables. The aim of this paper is to suggest, construct and evaluate such a procedure. 

We address this objective by amending standard DFM specifications by including a indicator variable for each factor loading, with a corresponding prior inclusion probability.  This constitutes a "sparse DFM", given that we purposely model sparsity in factor loadings. 

There are examples of Bayesian estimation of sparse static factor models \citep[see e.g.][]{LucasEtAl2006, CarvalhoEtAl2008}, inspired by Bayesian variable selection (BVS) \citep[see e.g. textbook by][]{TadesseVannucci2021}. \cite{KaufmannSchumacher2012} consider a similar approach allowing VAR-dynamics on the factors. \cite{KaufmannSchumacher2019} estimate sparse DFMs while dealing with identification by post-draw rotations. All of these methods use MCMC-techniques, attempting to sample from the posterior distribution. This is typically computationally heavy, which can make them unsuitable for forecasters working in fast-pace analytical environments. What if we assume computational constraint?

In this paper we develop an approximate solution to the sparse and large dynamic factor models, which can be updated frequently and fast on a standard computer, while also taking into account any data availability pattern.

We extend the method of \cite{Spanberg2021b} by introducing a, to our knowledge, novel VI approach for factor loading selection, also inspired by BVS. A common BVS technique is the spike-and-slab prior \citep[see e.g.][]{MitchellBeauchamp1988, GeorgeMcCulloch1993, IshwaranRao2005}. Generally speaking, the spike-and-slab prior is a density with a point mass (a spike) at a particular point and some other prior (a slab) elsewhere. To illustrate this idea, if we decompose some regressor parameter $\alpha = z\lambda$, where $ z \in \{0,1\}$ and $\lambda \in \mathbbm{R}$, a particular example of spike-and-slab prior is
\begin{align*}
    \lambda &\sim \mathcal{N}\left(0, v \right) \ \\
    z &\sim Bernoulli(\beta)
\end{align*}
where $\mathcal{N}(\cdot,\cdot)$ is the Gaussian distribution and $Bernoulli(\cdot)$ is the Bernoulli distribution. In other words, the prior for $\alpha$ is a point mass at 0 with probability $\beta$, and Gaussian with probability $(1 - \beta)$. 

The example above is an alternative to LASSO-regression \citep{Tibshirani1996} and other variable selecting regularisers, due to its sparsity inducing capabilities. 

In the spirit of this, we will employ a similar approach in a multivariate manner, explicitly modelling sparse solutions into the prior of factor loadings. By editing the algorithm of \cite{Spanberg2021b}, we can estimate the model in a computationally viable manner. Due to subsequent and frequent reference to their work, we will henceforth more conveniently refer to \cite{Spanberg2021b} as S21.

\section{Model framework}
Let $\{\ti{y}_t\}_{t=1}^T$ be a stochastic process, where $\ti{y}_t$ is a $n$-length column vector of time series at time $t$ with elements $\{\ti{y}_{i,t}, i=1, ..., n\}$, and $f_t$ be a $r$-length column vector of dynamic factors with elements $\{f_{j,t}, j=1, ..., r\}$, stacked in $F_t = [f_t' \; f_{t-1}' \; ... \; f_{t-p}']$. We also define $s=r(p+1)$, where $p$ is the number of loading lags. Our model can be specified in state space form:
\begin{align}
    \ti{y}_t &= \left(Z \circ \Lambda\right)F_t + \epsilon_t, \quad &\epsilon_t \sim \mathcal{N}\left(0, \Sigma_\epsilon \right), \label{yeqSS} \\
    F_t &= \widetilde{\Phi} F_{t-1} + S\ti{u}_t, \quad &\ti{u}_t \sim \mathcal{N}\left(0, \Sigma_\ti{u} \right), \label{feqSS}
\end{align}
where 
\begin{align*}
\widetilde{\Phi} &= \left[
    \begin{array}{cc}
      \multicolumn{2}{c}{\Phi}  \\ \hdashline[2pt/2pt]
       \multicolumn{1}{c}{\underset{rp \times rp}{I}} & \multicolumn{1}{;{2pt/2pt}c}{\underset{rp \times r}{0}}
    \end{array}\right], \quad S = \begin{bmatrix} I_r \\ 0_{rp \times r} \end{bmatrix}, \quad F_t = \begin{bmatrix} f_t \\ f_{t-1} \\ \vdots \\ f_{t-p} \end{bmatrix},  \end{align*}
and where we define parameters by
\begin{align*}
&\Lambda \text{ is a } (n \times s)\text{-matrix, with $i$th row } \lambda_i' \text{ and $(i,k)$-element } \lambda_{i,k} \in \mathbbm{R}, \text{ collecting factors loadings}; \\
&Z \text{ is a } (n \times s)\text{-matrix, with $i$th row }  z_i' \text{ and $(i,k)$-element }z_{i,k} \in \{0,1\}, \text{ collecting loading selectors};  \\
&\Phi \text{ is a } (r \times s)\text{-matrix, with $j$th row }  \phi_i' \text{ and $(j,k)$-element }\phi_{i,k} \in \mathbbm{R}, \text{ collecting transition parameters}; \\
&\Sigma_\epsilon \text{ is a } (n \times n)\text{-diagonal matrix, with $i$th diagonal element } \sigma^2_{\epsilon_i} \in \mathbbm{R}_{>0} \text{ collecting idiosyncratic variances;}\\
&\Sigma_\ti{u} \text{ is a } (r \times r)\text{-diagonal matrix, with $j$th diagonal element } \sigma^2_{\ti{u}_j} \in \mathbbm{R}_{>0} \text{ collecting factor residual variances.}
\end{align*} 
Furthermore, $(Z \circ \Lambda)$ is the Hadamard product (elementwise product) of $Z$ and $\Lambda$. The specification \eqref{yeqSS}-\eqref{feqSS} differs in essence from S21 by the inclusion of loading selectors $Z$.\footnote{It also differs in the more flexible inclusion of $\Sigma_{\ti{u}}$, whereas S21 restricts factor scaling by setting $\Sigma_{\ti{u}} = I_r$.} In other words, we have included binary on-and-off switches for each individual factor loading, which states are unknown and objects of estimation. The prior density of loading selectors as independent Bernoulli-distributed according to
\begin{align}
    z_{i,k} \sim Bernoulli\left(\beta_{i,k}\right), \quad i = 1, ..., n, \quad k = 1, ..., s, \label{zprior}
\end{align}

where $\beta_{i,k}$ consequently is the prior inclusion probability of $\lambda_{i,k}$. Practically, we have the option to use priors in order to direct the data analysis to some specific features. In the extreme end we can with 100 \% probability include or exclude factor loadings of our choosing, and dogmatically form structures. But we do not have to be dogmatic. The prior also allows for anything in between, directing to features with a chosen degree of certainty.

Beyond that, perhaps the most important benefit of the prior still remains even if we are impartial to certain structures (e.g. choosing all $\beta_{i,j}$ to be equal). The prior is sparsity inducing, meaning that it will lead to factor loading selection, setting some to zero.

Following S21 we define  $\{\Lambda, \Sigma_\epsilon\}$-priors\footnote{Our $\lambda_{i}$ priors are somewhat more flexible than S21, as they restrict shrinkage $V_{\lambda_i}$ to be equal for every $i$.}
\begin{align}
    &\lambda_i|\sigma^2_{\epsilon_i} \sim \mathcal{N}\left(0, \sigma_{\epsilon_i}^2 V_{\lambda_i}\right), \quad i=1, ..., n, \label{lambdaprior} \\
    &\sigma^2_{\epsilon_i} \sim \text{Scaled-Inv-}\chi^2\left(\nu_{\epsilon_i}, \tau^2_{\epsilon_i}\right), \quad i=1, ..., n. \label{sigmaeprior}
\end{align}

If factors were observed and loading selection was decided, then \eqref{lambdaprior}-\eqref{sigmaeprior} would be conjugate priors to observation equation \eqref{yeqSS} in accordance to ordinary Bayesian linear regression. $V_{\lambda_i}^{-1}$ determines the degree of shrinkage towards $\lambda_i = 0$, and $\nu_{\epsilon_i}$ determines the magnitude of prior information that $\sigma^2_{\epsilon_i} = \tau^2_{\epsilon_i}$. Equivalently we choose $\{\Phi, \Sigma_\ti{u} \}$-prior 
\begin{align*}
    &\phi_j|\sigma^2_{\ti{u}_j} \sim \mathcal{N}\left(0, \sigma_{\ti{u}_j}^2 V_{\phi_j}\right), \quad j = 1, ..., r  \\
    &\sigma^2_{\ti{u}_j} \sim \text{Scaled-Inv-}\chi^2\left(\nu_{\ti{u}_j}, \tau^2_{\ti{u}_j}\right), \quad j = 1, ..., r, 
\end{align*}
with the same interpretation for state equation \eqref{feqSS}. Lastly, we have a prior to initialize the dynamic factors according to
\begin{align*}
    F_0 \sim \mathcal{N}\left(0, V_{F_0}\right).
\end{align*}

It is worth pointing out that $V_{\lambda_i}$'s are not only affecting the degree of which factors are able to load onto the data, but also affects the scale of the factors themselves. Scaling of factors is arbitrary in the likelihood function as long as it follows by the opposite scaling of corresponding factor loadings. Even though the likelihood is unaffected by scale, the prior, and consequentely the posterior, is not. Prior shinkage on factor loadings will reflect onto different prior probabilities of scaling of factors as well. This influence is conjoined with parameters $V_{F_0}$, $\nu_{u_j}$ and $\tau^2_{u_j}$, which also provides prior information about factor scales.

\subsection{Estimation Algorithm} \label{sec.estalg} 

We define $\Omega$ as a set of all available observations in $\{ \ti{y}_t\}_{t=1}^T$, set of factors $F = \{F_t\}_{t=0}^T$ and a parameter set $\theta =  \{\Lambda, \Sigma_{\epsilon}, \Phi, \Sigma_{\ti{u}}\}$. Our goal is to find a \textit{variational density} $q(F,\theta,Z)$ which approximates posterior density $p\left(F,\theta,Z|\Omega\right)$. S21 uses the VI approach  Mean field approximation \citep[see e.g][]{BleiEtAl2017}, defining $q(\cdot)$ as belonging to a restricted class of densities with blockwise independent variables. Due to the blockwise nature of the restriction, this is called a structural mean field (SMF) approximation. 

In our case, we have to take into account loading selectors $Z$. By editing S21's approach with this added feature, we want to find a density 
\begin{align*}
    q(F,\theta,Z) = \underset{\tilde{q} \in \mathcal{Q}}{\text{arg min }} \text{KL}\left(\tilde{q}\left(F,\theta,Z\right) \Big|\Big|p\left(F,\theta,Z|\Omega\right)\right), \numberthis \label{qopt} 
\end{align*}
where $\mathcal{Q}$ is the set of all densities satisfying $\tilde{q}\left(F,\theta,Z\right) = \tilde{q}\left(F\right)\tilde{q}\left(\theta\right)\prod_{i=1}^n \prod_{k=1}^s \tilde{q}\left(z_{i,k}\right)$ and  $\text{KL}\left(\tilde{q}||p\right)$ is the \textit{Kullback-Leibler divergence} between $\tilde{q}$ and $p$ \citep{KullbackLeibler1951}. The general SMF solution for $q$ is known \citep[see e.g.][]{BleiEtAl2017}, which in our specific case translates into
\begin{align*}
    q(F) &\propto \exp\{\E{q(\theta,Z)}{\ln p\left(\Omega, F, \theta, Z \right)}\}, \\
    q(\theta) &\propto \exp\{\E{q(F,Z)}{\ln p\left(\Omega, F, \theta, Z \right)}\}, \\
    q(z_{i,k}) &\propto \exp\{\E{q(F,\theta,Z_{-(i,k)})}{\ln p\left(\Omega, F, \theta, Z \right)}\}, \quad  i=1, ..., n, \; k = 1, ..., s,
\end{align*}

where $Z_{-(i,k)}$ is the set of loading selectors, excluding $z_{i,k}$. As noted by S21, some parameter blocks share no common terms in the log joint density $\ln p\left(\Omega, F, \theta, Z \right)$. Consequently, we can decompose the variational density further according to \begin{align*}
    q(\theta) = \left(\prod_{i=1}^n q(\lambda_i|\sigma_{\epsilon_i}^2)q(\sigma_{\epsilon_i}^2)\right)\left(\prod_{j=1}^s q(\phi_j|\sigma_{\ti{u}_j})q(\sigma_{\ti{u}_j}^2)\right).
\end{align*}
To assess these densities, we define following first moments:
\begin{align*}
    \E{q(\Lambda|\Sigma_\epsilon)}{\Lambda} &\equiv M_\Lambda, \quad \E{q(\Sigma_\epsilon)}{\Sigma_\epsilon^{-1}} \equiv \Psi_\epsilon^{-1}, \quad \E{q(Z)}{Z} \equiv B,  \numberthis \\
     \E{q(\Phi|\Sigma_\ti{u})}{\Phi} &\equiv M_\Phi, \quad \E{q(\Sigma_\ti{u})}{\Sigma_\ti{u}^{-1}} \equiv \Psi_\ti{u}^{-1}. 
\end{align*}
We will also make use of some objects relating to second moments:
\begin{align*}
    P_i = \E{q(Z)}{z_i z_i'}, \quad Q_i = \sum_{t=1}^Ta_{i,t}\E{q(F)}{F_t F_t'}, \quad R_i = \E{q(\lambda_i|\sigma_{\epsilon_i}^2)}{\frac{\lambda_i \lambda_i'}{\sigma_{\epsilon_i}^2}}, \numberthis \label{PQRi}
\end{align*}
where $a_{i,t} = \mathbbm{1}\left\{\ti{y}_{i,t} \text{ is available}\right\}$ and $\mathbbm{1}\left\{\cdot\right\}$ is the indicator function. Additionally, like S21, we denote $T_i = \sum_{t=1}^T a_{i,t}$ as the number of available observations for variable $i$. 

Variational density of $\{\phi_j,\sigma_{\ti{u}_j}^2\}$ (derived in Appendix \ref{sec:A2}) is given by\footnote{These expressions are analogous with the results in S21. In S21 however, the factor residual variance is assumed to be known.}
\begin{align*}
    q(\phi_j|\sigma^2_{\ti{u}_j}) &= \mathcal{N}\left(\phi_j\Big|\mu_{\phi_j}, \sigma^2_{\ti{u}_j}\Sigma_{\phi_j} \right), \numberthis \label{qphi} \\
    q(\sigma^2_{\ti{u}_j}) &= \text{Scaled-Inv-}\chi^2\left(\sigma^2_{\ti{u}_j}\Big|\nu_{\ti{u}_j} + T, \psi^2_{\ti{u}_j}\right), \numberthis \label{qsigmau}
\end{align*}
where
\begin{align*}
       \mu_{\phi_j} &= \left(\sum_{t=1}^T\E{q(F)}{F_{t-1} F_{t-1}'} + V_{\phi_j}^{-1}\right)^{-1}\left(\sum_{t=1}^T\E{q(F)}{F_{t-1}f_{j,t}}\right), \numberthis \label{muphi} \\
       \Sigma_{\phi_j} &= \left(\sum_{t=1}^T\E{q(F)}{F_{t-1} F_{t-1}'} + V_{\phi_j}^{-1}\right)^{-1}, \numberthis \label{Sigmaphi} \\
       \psi_{\ti{u}_j}^2 &= \frac{1}{\nu_{\ti{u}_j} + T}\left(\nu_{\ti{u}_j} \tau^2_{\ti{u}_j} + \sum_{t=1}^T \E{q(F)}{f_{j,t}^2}- \mu_{\phi_j}'\Sigma_{\phi_j}^{-1}\mu_{\phi_j}\right). \numberthis \label{psiu} 
\end{align*}

 $\mu_{\ti{u}_j}'$ is the $j$th row of $M_\Phi$ and $\psi_{\ti{u}_j}^2$ is the $j$th diagonal element of $\Psi_{\ti{u}}$. Equations \eqref{qphi}-\eqref{psiu} take the form of a ordinary Bayesian linear regression with conjugate priors, but where the sufficient statistics are exchanged by "sufficient-like" statistics, given by sums of factor moments: $\sum_{t=1}^T\E{q(F)}{f_{j,t} F_{t-1}'}$, $\sum_{t=1}^T\E{q(F)}{F_{t-1} F_{t-1}'}$ and $\sum_{t=1}^T \E{q(F)}{f_{j,t}^2}$. If the factors were observed we could take away the expectation brackets, making these sufficient-like statistics exactly equal to sufficient statistics for a multivariate ordinary linear regression applied onto \eqref{feqSS}.

 Similarly, we can define analogous sufficient-like statistics for observation equation \eqref{yeqSS}. Define 
\begin{align*}
    g_i &= \sum_{t=1}^T \E{q(F)}{F_t} a_{i,t}  \ti{y}_{i,t}. \numberthis \label{gi}
\end{align*}
Then $g_i$, $Q_i$ and $\sum_{t=1}^2a_{i,t}\ti{y}^2_{i,t}$ are sufficient-like statistics for a linear regression of \eqref{yeqSS}, in the specific case where $Z = \mathbf{1}_{n \times s}$, with $\mathbf{1}_{n \times s}$ being a matrix of ones. These are fundamental objects in S21's variational density of $\{\lambda_i, \sigma^2_{\epsilon_i}\}$.

We take into account an unknown $Z$, giving the variational density (derived in Appendix \ref{sec:A1}): 
 \begin{align*}
     q(\lambda_i | \sigma^2_{\epsilon_i}) &= \mathcal{N}\left(\lambda_i\Big| \mu_{\lambda_i} , \sigma_{\epsilon_i}^2 \right), \\
     q(\sigma^2_{\epsilon_i}) &= \text{Scaled-Inv-}\chi^2\left(\sigma^2_{\epsilon_i}\Big| \nu_{\epsilon_i} + T_i, \psi_\ti{u}^2 \right),
\end{align*}
where
\begin{align*}
       \mu_{\lambda_i} &= \left(P_i \circ Q_i + V_{\lambda_i}^{-1}\right)^{-1}\left(b_i \circ g_i \right), \numberthis \label{mulambdai} \\
       \Sigma_{\lambda_i} &= \left(P_i \circ Q_i + V_{\lambda_i}^{-1}\right)^{-1}, \numberthis \label{Sigmalambdai} \\
       \psi_{\epsilon_i}^2 &= \frac{1}{\nu_{\epsilon_i} + T_i}\left(\nu_{\epsilon_i}\tau^2_{\epsilon_i} + \sum_{t=1}^T a_{i,t}\ti{y}_{i,t}^2 - \mu_{\lambda_i}'\Sigma_{\lambda_i}^{-1}\mu_{\lambda_i}\right), \numberthis \label{psiepsiloni} 
\end{align*}
and where $b_i'$ is the $i$th row of $B$.

Expressions \eqref{mulambdai}-\eqref{psiepsiloni} also have some similarities to Bayesian linear regression, but where sufficient-like statistics $g_i$ and $Q_i$ are weighted by first moment ($b_i$) and second moment ($P_i$) of $z_i$, respectively. 

The expressions give us an intuitive interpretation of how the variational density is updated by data. Our information about $\lambda_i$ is only updated by available observations and only to a degree determined by its inclusion probability. More specifically, $g_i$, $Q_i$ and $\sum_{t=1}^T a_{i,t} \ti{y}_{i,t}^2$ are sums corresponding to available elements in variable $i$. If there are no available elements, \eqref{mulambdai}-\eqref{psiepsiloni} reduce to prior parameters ($0, V_{\lambda_i}$ and $\tau^2_{\epsilon_i}$, respectively). In a similar manner, updating the density of $\lambda_i$ is contingent upon, and weighted by, posterior inclusion probability. If $b_i = 0$ then $P_i = 0$ (as we soon will show), meaning again that the variational density of $\lambda_i$ reduces to its prior. 

By knowing expressions for the moments of $\lambda_i$, we can deduce an explicit expression for $R_i$:
\begin{align*}
    R_i = \Sigma_{\lambda_i} + \frac{1}{\psi_{\epsilon_i}^2} \mu_{\lambda_i} \mu_{\lambda_i}', \numberthis \label{Ri}
\end{align*}
which is an object used in the variational density of $Z$. The variational densities of individual elements in $Z$ (derived in Appendix \ref{sec.A4}) are given by
\begin{align*}
    q(z_{i,k}) = Bernoulli\left(z_{i,k}\Big|b_{i,k}\right), i = 1, .., n, k = 1, ..., s,
\end{align*}
where 
\begin{align*}
    b_{i,k} &= \text{expit}\left(\gamma_{i,k} + \text{logit}\left(\beta_{i,k}\right) \right), \numberthis \label{bik} \\
    \gamma_{i,k} &= \frac{\mu_{\lambda_{i,k}}g_{i,k}}{\psi_{\epsilon_i}^2} - \frac{1}{2}[R_i]_{k,k} [Q_i]_{k,k} - \sum_{m \neq k} b_{i,m} [R_i]_{k,m}[Q_i]_{k,m}.
\end{align*}
$b_{i,k}, \mu_{\lambda_{i,k}}$ and $g_{i,k}$  are the $k$th elements of $b_i, \mu_{\lambda_i}$ and $g_i$, respectively. Additionally, we use functions $\text{expit}\left(c\right) = 1/(1 + e^{-c})$ and $\text{logit}\left(d\right) = \ln\left(d/(1-d)\right)$. 

All non-prior about $z_{i,k}$ is contained in the statistic $\gamma_{i,k}$; if there are no available observations of variable $i$, then $\gamma_{i,k} = 0$ and consequently $b_{i,k} = \beta_{i,k}$. The larger $\gamma_{i,k}$ is, the higher posterior inclusion probability.

By studying its terms we get some intuitions. The first term adds to $\gamma_{i,k}$ by a scaled size of the first moment of $\lambda_{i,k}$, relative to the idiosyncratic variance. The remaining two terms, on the other hand, reduces $\gamma_{i,k}$ by scaling the second moment of $\lambda_{i,k}$ (again relative to idiosyncratic variance) by the availability-sum of the second moments of the regressor $F_{t,k}$ and by bilateral product moments and other loadings' inclusion probabilities. 

Interpreted rather simplified, inclusion of $\lambda_{i,k}$ is more probable if: 
\begin{itemize}
    \item[1.] its mean estimate is large,
    \item[2.] its variance is small,
    \item[3.] there are not too many other large factor loadings with high inclusion probability and
    \item[4.] it does not correlate too much with other factor loadings with high inclusion probability.
\end{itemize}  In a sense, the factor loadings corresponding to a specific variable are "stealing" inclusion probability from each other, where one strong inclusion pushes another towards non-inclusion, especially if they have large bilateral covariance. In other words, if several factor loadings describe almost the same thing, this approach has a tendency to bring forward only a few or even just one of them. The researcher can, if they like, give prior credence to specific loadings by differing $\beta_{i,k}$'s. 

As $z_{i,k}$ is Bernoulli-distributed, we can get a explicit expression for $P_i$:
\begin{align*}
    [P_i]_{k,m} = \begin{cases} b_{i,k} &\text{ if } k=m \\
                                b_{i,k}b_{i,m} &\text{ if } k\neq m \end{cases}, \numberthis \label{Pi}
\end{align*}
where we have used the variational restriction that all elements in $Z$ are independent. 

Lastly, we need an expression for $q(F)$. Define $\underset{Z \circ \Lambda}{M} \equiv  \E{q(\theta,Z)}{Z \circ \Lambda} = B \circ M_\Lambda $. Again, by extending the results of S21 (derived in Appendix \ref{sec:A3}) we get
\begin{align*}
   q(F) &= C_F\exp\left\{\frac{1}{2}\sum_{t=1}^T\left(\begin{bmatrix} \ti{y}_t \\ 0_{s \times 1} \end{bmatrix} - \begin{bmatrix} \underset{Z \circ \Lambda}{M} \\[7pt] I_{s} \end{bmatrix}  F_t\right)'\begin{bmatrix}A_t\Psi_\epsilon^{-1} & 0_{n \times s} \\ 0_{s \times n} & \Sigma^{\theta Z}_t \end{bmatrix}\left(\begin{bmatrix} \ti{y}_t \\ 0_{s \times 1} \end{bmatrix} - \begin{bmatrix} \underset{Z \circ \Lambda}{M} \\[7pt] I_{s} \end{bmatrix}  F_t\right) \right. \\
    &\qquad \left. - \frac{1}{2}\sum_{t=1}^T \left(F_t - \widetilde{M}_\Phi F_{t-1}\right)'\widetilde{\Psi}_\ti{u}^{-1}\left(F_t - \widetilde{M}_\Phi F_{t-1}\right) -  \frac{1}{2}F_0'\left(V_{F_0}^{-1} + \sum_{j=1}^r \Sigma_{\phi_j}\right)F_0 \right\}, \numberthis \label{q_F}
\end{align*}
which describes state space system 
\begin{align}
    \widetilde{\ti{y}}_t &= \underset{Z \circ \Lambda}{\widetilde{M}} F_t + \widetilde{\epsilon}_t, \quad \widetilde{\epsilon}_t \sim \mathcal{N}\left(0, \widetilde{\Sigma}_t \right), \label{Fss1} \\ 
    F_t &= \widetilde{M}_\Phi F_{t-1} + S\ti{u}_t, \quad \ti{u}_t \sim \mathcal{N}\left(0, \Psi_\ti{u}\right), \label{Fss2} \\
    F_0 &\sim \mathcal{N}\left(0, \Sigma_{F_0}\right), \label{Fss3}
\end{align}
where
\begin{align*}
    &\widetilde{\ti{y}}_t = \begin{bmatrix} \ti{y}_t \\ 0_{s \times 1} \end{bmatrix}, \quad \underset{Z \circ \Lambda}{\widetilde{M}} = \begin{bmatrix} \underset{Z \circ \Lambda}{M} \\[7pt] I_{s} \end{bmatrix}, \quad \widetilde{M}_\Phi = \left[
    \begin{array}{cc}
      \multicolumn{2}{c}{M_\Phi}  \\ \hdashline[2pt/2pt]
       \multicolumn{1}{c}{\underset{rp \times rp}{I}} & \multicolumn{1}{;{2pt/2pt}c}{\underset{rp \times r}{0}}
    \end{array}\right], \quad  \widetilde{\Sigma}_t =  \begin{bmatrix} \Psi_\epsilon & 0_{n \times s} \\ 0_{s \times n} & \left(\Sigma^{\theta Z}_t\right)^{-1}  \end{bmatrix}, \\
    &\Sigma_{F_0} = \left(V_{F_0}^{-1} + \sum_{j=1}^r \Sigma_{\phi_j}\right)^{-1}, \\
     &\Sigma^{\theta Z}_t = \begin{cases}  \sum_{i=1}^n a_{i,t}\Big(\left(P_i \circ \Sigma_{\lambda_i}\right) +  \text{diag}\left(w_i\right)\Big) + \sum_{j=1}^r \Sigma_{\phi_j}\quad &\text{ if } t=1, ..., T-1, \\ \\
    \sum_{i=1}^n a_{i,t}\Big(\left(P_i \circ \Sigma_{\lambda_i}\right) +  \text{diag}\left(w_i\right)\Big) \quad &\text{ if } t=T, \end{cases}
\end{align*}
and where $w_i'$ is the $i$th row of
\begin{align*}
    W =  \Psi_\epsilon^{-1}(B \circ (\mathbf{1}_{n \times s} - B) \circ M_\Lambda \circ M_\Lambda).
\end{align*}

The variational density of $F$ is thus given by the state density obtained by running Kalman filter and smoother over a DFM with parameters $\{\underset{Z \circ \Lambda}{M}, M_\Phi, \Psi_\epsilon, \Psi_\ti{u}\}$, but where the observation vector is augmented by zeros, corresponding to an additional residual variance $\Sigma_t^{\theta Z}$. The latter takes into account parameter uncertainty. 

Some essentials are very similar to S21, but it differs in some core aspects. First, our factor loadings are weighted by inclusion probability $B$. Second, our expression of $\Sigma^{\theta Z}_t$ takes into account the uncertainty of $Z$, effectively by weighting and amending the covariance of factor loadings. Similar to expressions for $q(\Lambda,\Sigma_\epsilon)$, $Z$ comes in by weighting core objects in the system. In the specific case were all loadings have 100 \% inclusion probability, it reduces in essence to S21's result. 

Using S21's collapsing of the observation vector \citep[see][for the general solution]{JungbackerKoopman2015}, we can exchange the observation equation \eqref{Fss1} by the potentially much faster
\begin{align*}
    \ti{y}^\star_t &=  F_t + \epsilon^\star_t, \quad \epsilon^\star_t \sim \mathcal{N}\left(0, H_t^\star \right), \numberthis \label{FssC}
\end{align*}
where 
\begin{align*}
   \ti{y}^\star_t &=  \left(\underset{Z \circ \Lambda}{M}'A_t\Psi_\epsilon^{-1} \underset{Z \circ \Lambda}{M} + \Sigma^{\theta Z}_t \right)^{-1}\underset{Z \circ \Lambda}{M}'\Psi_\epsilon^{-1}A_t\ti{y}_t, \\
   H^\star_t &= \left(\underset{Z \circ \Lambda}{M}'A_t\Psi_\epsilon^{-1} \underset{Z \circ \Lambda}{M} + \Sigma^{\theta Z}_t \right)^{-1}.
\end{align*}
This reduces the length of the observation vector from $n$ to $s$. By running the Kalman filter and smoother over \eqref{FssC} together with \eqref{Fss2}-\eqref{Fss3}, we can obtain desired moments of $F_t$ and related objects, including $Q_i$. 

We have shown that the individual variational densities $q(\Lambda,\Sigma_\epsilon)$, $q(\Phi, \Sigma_u)$, $q(Z)$ and $q(F)$ each can be summarized by a number of parameters. These parameters are mostly non-linear functions of each other, which impedes a clear direct solution. To find a local optimal solution to the optimization problem \eqref{qopt}, we update the parameters recursively, conditioned on parameters (and thus densities) from previous iterations. Algorithm \ref{algoVIDFM} describes the procedure. 

As minimizing KL-divergence is equal to maximizing the Evidence Lower Bound (ELBO) \citep[see e.g.][]{BleiEtAl2017}, a reasonable convergence criteria is checking if relative changes in ELBO between iterations is below some tolerance. ELBO is provided (and derived) in Appendix \ref{sec:A5}.

\begin{algorithm} 
  \caption{VI for DFM with loading selection prior} \label{algoVIDFM}
  \begin{itemize}
    \item[\textbf{1.}] \textbf{Initial settings} 
    \begin{itemize}
     \item[a)] Set numbers of factors $r$, lag-length $p$, and compute $s=r(p+1)$.
     \item[b)] Set prior parameters \begin{itemize} 
     \item[$\bullet$] $V_{F_0}$,
     \item[$\bullet$] $V_{\lambda_i}$, $\nu_{\epsilon_i}$ and $\tau^2_{\epsilon_i}$, $\forall i = 1, ..., n$,
     \item[$\bullet$] $V_{\phi_j}$, $\nu_{u_j}$ and $\tau^2_{u_j}$, $\forall j=1,..,r$,
     \item[$\bullet$] and  $\beta_{i,k}$, $\forall i = 1, ..., n, k=1, ..., s.$
     \end{itemize} 
          \item[c)] Set starting parameters
     \begin{itemize} 
     \item[$\bullet$] $M_\Lambda^{(0)}$, $M_\Phi^{(0)}$, $\Psi_\epsilon^{(0)}$, $\Psi_u^{(0)}$, $B^{(0)}$, 
     \item[$\bullet$] $\Sigma_{\lambda_i}^{(0)}, \forall i=1, ..., n,$
     \item[$\bullet$] $\Sigma_{\phi_j}^{(0)}, \forall j=1,...,r$. 
     \item[$\bullet$] and compute $P_i^{(0)}$, $\forall i = 1, ..., n$, from \eqref{Pi} using $B^{(0)}$. \end{itemize}
     \item[d)] Set tolerance level. 

     \item[e)] Set starting index $d \leftarrow 0$.
    \end{itemize}
    \item[\textbf{2.}] \textbf{Update variational densities}
    
    \textbf{while} criteria $>$ tolerance level 
        \begin{itemize}
        \item[a)] $d \leftarrow d + 1$
        \item[b)] Update $q_d(F)$: 
        \begin{itemize} \item[$\bullet$] Run model \eqref{FssC} with \eqref{Fss2}-\eqref{Fss3}, using parameters $M_\Lambda^{(d-1)}$, $M_\Phi^{(d-1)}$, $\Psi_\epsilon^{(d-1)}$, $\Psi_u^{(d-1)}$, $B^{(d-1)}$, $\Sigma_{\lambda_i}^{(d-1)}, \forall i=1, ..., n,$ and  $\Sigma_{\phi_j}^{(d-1)}, \forall j=1,...,r$. through the Kalman filter and smoother. 
         \item[$\bullet$] Compute $\sum_{t=1}^T \E{q_d(F)}{F_{t-1} F_{t-1}'}$, $\E{q_d(F)}{f_{j,t} F_{t-1}'}$ and $\E{q_d(F)}{f_{j,t}^2}$, $\forall j = 1, ..., r$, using Kalman smoother results. 
        \item[$\bullet$] Compute $Q_i^{(d)}$ from \eqref{PQRi} and  $g_i^{(d)}$ from \eqref{gi}, $\forall i = 1, ..., n$,  using Kalman smoother results. 
\end{itemize}
        \item[c)] Update $q_d(\Phi, \Sigma_u)$: 
        \begin{itemize} 
        \item[$\bullet$] Compute $M_\Phi^{(d)}$, $\Psi_u^{(d)}$ and $\Sigma_{\phi_j}^{(d)}$ from \eqref{muphi}-\eqref{psiu} using $\sum_{t=1}^T \E{q_d(F)}{F_{t-1} F_{t-1}'}$, $\E{q_d(F)}{f_{j,t} F_{t-1}'}$ and $\E{q_d(F)}{f_{j,t}^2}$, $\forall j = 1, ..., r$. 
        \end{itemize}
        \item[d)] Update $q_d(\Lambda, \Sigma_\epsilon)$: 
        \begin{itemize} 
        \item[$\bullet$] Compute $M_\Lambda^{(d)}$, $\Psi_\epsilon^{(d)}$ and $\Sigma_{\lambda_i}^{(d)}$ from \eqref{mulambdai}-\eqref{psiepsiloni} using $Q_i^{(d)}$and $g_i^{(d)}$, $\forall i = 1, ..., n$.
        \item[$\bullet$] Compute $R_i^{(d)}$ from \eqref{Ri} using $M_\Lambda^{(d)}$, $\Psi_\epsilon^{(d)}$ and  $\Sigma_{\lambda_i}^{(d)}$, $\forall i = 1, ..., n$.
        \end{itemize}
        \item[e)] Update $q_d(Z)$:
        \begin{itemize}
        \item[$\bullet$] \textbf{for} $i = 1, ..., n$
        
        \: \textbf{for} $k = 1, ..., s$
        
        \: \:    Compute $b_{i,k}^{(d)}$ from \eqref{bik} using $M_\Lambda^{(d)}$, $\Psi_\epsilon^{(d)}$, $R_i^{(d)}$,  $Q_i^{(d)}$, $g_i^{(d)}$, $b_{i,m}^{(d)}$, $\forall m = 1, ..., k-1$, and $b_{i,\ell}^{(d-1)}$, 
        
        \: \: $\forall \ell = k+1, ..., s.$ 
        
        \: \textbf{end for} 
            
        \textbf{end for}
        \item[$\bullet$] Compute $P_i^{(d)}$ from \eqref{Pi} using $B^{(d)}$
        \end{itemize}
        \end{itemize}
        \textbf{end while} 
  \end{itemize}
\end{algorithm}

\begin{algorithm} 
  \caption{Rerunning VI-DFM using initial inclusion} \label{algoRerunVIDFM}
  \begin{itemize}
    \item[\textbf{1.}] \textbf{Initial run} 
    \begin{itemize}
     \item[a)] Specify initial settings  in Algorithm 1 (Steps 1a-1e), with particular choice $B^{(0)} = \mathbf{1}_{n \times s}$.
     \item[b)] Update densities in Algorithm 1 (Steps 2a-2e). 
     \item[c)] Denote $\theta^\star$ as the parameter set of $M_\Lambda^{(D)}$, $M_\Phi^{(D)}$, $\Psi_\epsilon^{(D)}$, $\Psi_u^{(D)}$, $\Sigma_{\lambda_i}^{(D)}$, $\forall i=1, ..., n, \Sigma_{\phi_j}^{(D)}$, $\forall j=1,...,r$, where $D$ is the last iteration of Algorithm 1. 
     \end{itemize}
     \item[\textbf{2.}] \textbf{Rerunning}
     
      \textbf{while} $\text{comparison criteria} > 0$
        \begin{itemize} 
            \item[a)] Re-specify initial settings in Algorithm 1 (Steps 1a-1e) with starting parameters $\theta^\star$ and $B^{(0)} = \mathbf{1}_{n \times s}$. 
            \item[b)] Re-update densities in Algorithm 1 (Steps 2a-2e).
            \item[c)] Reset $\theta^\star$ as the parameter set of $M_\Lambda^{(D)}$, $M_\Phi^{(D)}$, $\Psi_\epsilon^{(D)}$, $\Psi_u^{(D)}$, $\Sigma_{\lambda_i}^{(D)}$, $\forall i=1, ..., n, \Sigma_{\phi_j}^{(D)}$, $\forall j=1,...,r$, where $D$ is the last iteration of the re-run of Algorithm 1. 
        \end{itemize} 
      \textbf{end while}
    \end{itemize}
\end{algorithm}

It is worth noting that the algorithm converges to a local optimum; it is inclined to converge to a particular identification of the factor model, with a particular sparsity pattern. The researcher might want to examine the optimization space by testing different starting values. One suggestion by S21 is to re-run the algorithm using rotations of the previous result as new starting values. If there are no identifying restrictions \citep[see][]{BaiWang2014}, that search space can be infinite. Still, there might be value in investigating well considered rotations. 

We complement their suggestion with another.  There is a risk that a factor loading inclusion probability is wrongly set close to zero in an early recursion and getting stuck there. To give factor loadings a second chance, we can re-run the algorithm with full inclusion as starting parameter. More concretely, the suggestion is to set the starting parameters equal to the previous end results, except loading inclusion probabilities, $b_{i,k}^{(0)}$, which are all set to 1. This will add parameter uncertainty of the previously non-included loading in the first factor estimation iteration, with a chance of providing new results. The procedure can be iterated until the re-run does not improve some evaluating criteria (e.g. ELBO). The re-running suggestion is described in Algorithm \ref{algoRerunVIDFM}. 

\section{Monte Carlo simulation study}

In this section we evaluate the estimation algorithm provided in Section \ref{sec.estalg}, using two seperate Monte Carlo simulation experiments. In the first experiment we simulate several seperate data sets under different DFM specifications, and evaluate loading inclusions, loading estimation errors and factor space estimation precisions. They are also compared to ML-estimation. 

In the second experiment we simulate from two larger DFMs, with different missing data patterns, and deep dive into estimation results and sparsity patterns for those particular simulations. 

In parts comparable, but not identical, to \cite{DozEtAl2011}, \cite{BanburaetAl2014} and \cite{SolbergerSpanberg2020} \citep[see also][]{Spanberg2021a}, we simulate data and factors by
\begin{align*}
        \ti{y}_t &= \Lambda F_t + \epsilon_t, \quad &\epsilon_t &\sim \mathcal{N}\left(0, \Sigma_\epsilon \right), \numberthis \label{sim1} \\
    f_t &= \Phi_1 f_{t-1} + \ti{u}_t, \quad &\ti{u}_t &\sim \mathcal{N}\left(0, I_r \right), \numberthis \\
    F_t &= [f_t' \; f_{t-1}' \; ... f_{t-p-1}']', \quad &t&=1, ..., T, \numberthis
\end{align*}
where 
\begin{align*}
     \begin{cases}[\text{vec}(\Lambda)]_\gamma  \sim \mathcal{N}(0,1) &\text{if }  \gamma \in \mathfrak{B} \\ [\text{vec}(\Lambda)]_\gamma = 0 & \text{if } \gamma \notin \mathfrak{B} \\  \end{cases}, \quad \gamma = 1, 2, ..., n\times s, \numberthis 
\end{align*}
$\mathfrak{B}$ is a $\lfloor \omega \times n \times s \rceil$-length vector of index-elements, provided by uniform random samples without replacement from  $[1 \; 2 \; ... \; s \times n]$, where $\lfloor C\rceil$ is a function rounding $C$ to nearest integer, and 
\begin{align*}
    \alpha_j &\sim U(-0.95,0.95), \quad \quad  &\xi_i &\sim U(0.1, 0.9), \numberthis \\ 
    [\Phi_1]_{j,m} &=\begin{cases} \alpha_j &\text{if } j = m, \\
     0 &\text{if } j \neq m, \end{cases} \quad \quad &[\Sigma_\epsilon]_{i,i} &= \begin{cases} \frac{\xi_i}{1-\xi_i}\zeta_i &\text{if } \zeta_i > 0, \\
    1 &\text{if } \zeta_i = 0, \end{cases}  \numberthis \\
    &         &\zeta_i &= \sum_{\ell=0}^p \sum_{j=1}^r \frac{1}{1-\alpha_j^2}[\Lambda]_{i,\ell r+j}^2,  \numberthis \label{simlast}
\end{align*} for $ i = 1, ..., n, \; j = 1, ..., r, \; m = 1, ..., r, \; k = 1, ..., s$.

Number of factors $r$, sample size $T$ and lag-length $p$ are controlled by the researcher, together with one additional parameter: $\omega \in [0 \; 1]$, which denotes the share of factor loadings included. The rest of the system is simulated. 

The included factor loadings are chosen by uniform random sampling without replacement. The \textit{signal-to-noise ratio}, $\text{Var}(\epsilon_{i,t}/y_{i,t})$, given by $\xi_i$, governs how much of individual variable variance is explained by the idiosyncratic component. In the trivial case with no included loadings, all variance is in the idiosyncratic component. The factors are simulated by autoregressive processes with one lag, where $\alpha_j$ is the $j$th factor's corresponding autoregressive parameter. To summarize, this system simulate factors with different persistence, loading onto variables differently with randomly chosen loadings with random magnitude, and random signal-to-noise ratios. 

\subsection{Simulation experiment 1} \label{sec.sim1} 

In the first experiment we evaluate the method in three regards: correct categorizing of loading inclusion, factor loading estimation errors and factor estimation precision. Factors, factor loadings and data are simulated using \eqref{sim1}-\eqref{simlast}. The simulations are done with different specifications in some categories: \begin{itemize}
\item number of variables $n$  as 50, 100 and 400,
\item sample sizes $T$ as 100 and 200,
\item number of factors $r$ as 1 and 2,
\item lag-length $p$ as 0 and 2,
\item loading inclusion share $\omega$ as 20 \%, 50 \% and 100 \%. 
\end{itemize} 

We run 200 simulations for each combination of specifications. After each simulation we standardize observation variables to unit standard deviation and estimate three models using Algorithm \ref{algoRerunVIDFM}, with different prior inclusion probabilities 20 \%, 50\% and 100 \%, respectively, for $\beta_{i,k}, \forall i,k$. Additionally, we estimate models with ML-estimation with no loading selection for comparison, using Expectation Maximization, similar to \cite{BanburaetAl2014}.\footnote{Our approach differs in that we have integrated out missing data preemptively. \cite{BanburaetAl2014} do not, which makes their estimator unnecessary complicated, as noted by \cite{Spanberg2021a}. The maximum likelihood should in principle be the same.}

There are no identification restrictions in the estimation. The procedure is consequently dependent on its capability to identify the true system. We help system-identification only in one singular aspect: we provide post-estimation is column and sign regulation, as to correctly compare the true factors and loadings to their estimated counterpart. For example, if the estimated second factor correlates higher with the true first factor, we allow a column change in the factor space and factor loading matrix, and if the correlation is negative we make the corresponding sign changes. We do not, however, make any rotations beyond that, which could effect the inclusion pattern. 

Variational inference with loading selection priors will subsequently be called VI-LS. Beyond $\beta$, the same priors are used for each VI-LS estimation. These are heuristically chosen, with no deeper procedure than basic reasoning. We use the prior knowledge of unit standard deviation, due to standardization, as a reference point in prior selection. The prior hyperparameters are chosen as:
\begin{itemize}
    \item The prior degrees of freedom $\nu^2_{\epsilon_i}$ and $\nu^2_{u_j}$ are selected as 1 $\forall i = 1, ..., n$ and $j = 1, ..., r$.
    \item The prior variance scale $\tau^2_{\epsilon_i}$ and $\tau^2_{u_j}$ are selected as 1 $\forall i = 1, ..., n$ and $j= 1, ..., r$.
    \item Shrinkage matrices $V_{F_0}$, $V_{\lambda_i}$ and $V_{\phi_j}$ are selected as diagonal matrices with 2 as each diagonal element, $\forall i = 1, ..., n$ and $j = 1, ..., r$.
\end{itemize}

In other words, we believe a-priori that variables and factors are white noise processes, with one prior observation each that they have standard deviation 1. There is some, but not very hard, shrinkage towards these white noise processes, with each individual loading having twice the amount of prior variance as the variance in data itself. 

Lastly, we decide the starting values for the algorithm: For ML we use results from standard linear regressions upon principal component estimated factors, and for VI-LS we use results from standard Bayesian linear regression upon ML-estimated factors.

In the first evaluation we say that the model has correctly categorized the inclusion of a factor loading if the estimated posterior probability inclusion, $b_{i,k}$, is above 50 \% for a true inclusion and below or equal to 50 \% for a true non-inclusion. For each given simulation and choice of $\beta$, we compute the statistic
\begin{align*}
    P_{Z} = \frac{1}{ns}\sum_{i=1}^n \sum_{k=1}^s \mathbbm{1}\left\{z_{i,k}^\star = \mathbbm{1}\{b_{i,k} > 0.5\} \right\}, \numberthis \label{PZ}
\end{align*} 

which measures the share of correctly selected loadings. $z^\star_t$ is an indicator which is 1 if $\lambda_{i,t}$ is truly included, and 0 otherwise. When using prior inclusion probability 100 \%, we already know that $P_Z = \omega$ and the same goes for ML; these estimates are conveniently left out from the table. 

Table \ref{T:sim1LSshare} shows the average $P_Z$ over simulations per specification scheme. 

\begin{table}[!ht]
\caption{Average share of correctly VI-LS estimated factor loading inclusions, in percent (Experiment 1)}
\label{T:sim1LSshare}
\centering 
\renewcommand{\arraystretch}{1.1}
\resizebox{\textwidth}{!}{
\begin{tabular}{l cc r ccr ccr ccr cc}
\cmidrule{4-15}										&		&		&		&	\multicolumn{2}{c}{$r=1,\;,p=0$}	&		&	\multicolumn{2}{c}{$r=1,\; p=2$}	&		&	\multicolumn{2}{c}{$r=2,\;p=0$}		&		&	\multicolumn{2}{c}{$r=2,\; p=2$}	\\	\cmidrule{5-6}	\cmidrule{8-9}	\cmidrule{11-12}	\cmidrule{14-15}
$\omega$	&	$n$	&	$T$	&	$\beta:$	&	20\%	&	50\%	&&	20\%	&	50\%	&&	20\%	&	50\%	&&	20\%	&	50\%	\\	\hline		
\multirow{6}{*}{20\%}	&	50	&	100	&		&	\textbf{97.16}	&	86.38	&&	\textbf{90.65}	&	80.98	&&	\textbf{96.16}	&	84.73	&&	\textbf{89.74}	&	79.88	\\			
	&	50	&	200	&		&	\textbf{97.46}	&	87.06	&&	\textbf{90.62}	&	80.41	&&	\textbf{96.22}	&	85.72	&&	\textbf{91.23}	&	80.99	\\			
	&	100	&	100	&		&	\textbf{97.03}	&	86.43	&&	\textbf{90.88}	&	80.70	&&	\textbf{96.12}	&	84.93	&&	\textbf{91.50}	&	81.13	\\			
	&	100	&	200	&		&	\textbf{97.47}	&	86.11	&&	\textbf{92.55}	&	81.98	&&	\textbf{96.27}	&	84.85	&&	\textbf{91.66}	&	81.30	\\			
	&	400	&	100	&		&	\textbf{97.30}	&	86.19	&&	\textbf{92.61}	&	81.82	&&	\textbf{96.42}	&	84.37	&&	\textbf{92.79}	&	82.06	\\			
	&	400	&	200	&		&	\textbf{97.46}	&	86.74	&&	\textbf{93.75}	&	83.14	&&	\textbf{96.02}	&	83.90	&&	\textbf{93.48}	&	82.80	\\	\\		
\multirow{6}{*}{50\%}	&	50	&	100	&		&	\textbf{97.96}	&	91.57	&&	\textbf{82.32}	&	79.07	&&	\textbf{90.03}	&	84.29	&&	\textbf{76.92}	&	75.65	\\			
	&	50	&	200	&		&	\textbf{98.46}	&	91.92	&&	\textbf{84.27}	&	80.45	&&	\textbf{91.09}	&	85.57	&&	\textbf{80.82}	&	77.68	\\			
	&	100	&	100	&		&	\textbf{97.84}	&	91.56	&&	\textbf{84.84}	&	81.42	&&	\textbf{90.76}	&	84.59	&&	\textbf{79.88}	&	77.93	\\			
	&	100	&	200	&		&	\textbf{98.38}	&	91.66	&&	\textbf{88.22}	&	83.38	&&	\textbf{90.88}	&	83.94	&&	\textbf{83.39}	&	79.29	\\			
	&	400	&	100	&		&	\textbf{97.92}	&	91.39	&&	\textbf{86.10}	&	82.21	&&	\textbf{90.82}	&	83.66	&&	\textbf{80.74}	&	78.12	\\			
	&	400	&	200	&		&	\textbf{98.44}	&	91.81	&&	\textbf{89.19}	&	84.17	&&	\textbf{89.85}	&	82.93	&&	\textbf{84.43}	&	80.33	\\	\\		
\multirow{6}{*}{100\%}	&	50	&	100	&		&	99.28	&	\textbf{99.91}	&&	67.01	&	\textbf{78.63}	&&	78.74	&	\textbf{85.89}	&&	49.97	&	\textbf{66.16}	\\			
	&	50	&	200	&		&	99.97	&	\textbf{100.00}	&&	76.57	&	\textbf{84.77}	&&	85.56	&	\textbf{90.09}	&&	61.14	&	\textbf{73.96}	\\			
	&	100	&	100	&		&	99.22	&	\textbf{99.83}	&&	66.94	&	\textbf{78.74}	&&	79.69	&	\textbf{86.81}	&&	51.88	&	\textbf{68.45}	\\			
	&	100	&	200	&		&	99.97	&	\textbf{100.00}	&&	77.33	&	\textbf{85.69}	&&	86.02	&	\textbf{90.60}	&&	63.63	&	\textbf{75.94}	\\			
	&	400	&	100	&		&	99.32	&	\textbf{99.88}	&&	69.28	&	\textbf{80.83}	&&	80.82	&	\textbf{87.75}	&&	52.88	&	\textbf{69.38}	\\			
	&	400	&	200	&		&	99.98	&	\textbf{100.00}	&&	78.29	&	\textbf{86.47}	&&	87.05	&	\textbf{91.61}	&&	64.47	&	\textbf{77.02}	\\ 	\hline		
\end{tabular}}
    \begin{minipage}{\textwidth}
    \small 
    \emph{Note: Bold-face denotes first place (or shared first place) in evaluation criteria. $n$ is number of variables, $T$ is sample size, $r$ is number of factors, $p$ is number of factor lags, $\omega$ is share of included factor loadings in simulation, $\beta$ is prior probability of inclusion per factor loading. }
    \end{minipage}
\end{table}

The estimations have the best track record when there is no loading lag-length. In the smallest case ($r=1$ and $p=0$) we see a very large share of correctly included loadings, especially for prior $\beta_{i,k} = 0.2$. Encouragingly in this case, the procedure almost select every loading when $\omega = 1$ as well, performing very close to the correct specification. Very small, but truly included, loadings are the hardest to categorize. Luckily, those are the least important, as they have relatively negligible effects on the variables regardless of inclusion or not.

It is somewhat harder to correctly distinguish factor loadings when having loading lags, especially those corresponding to highly persistent factors. This is partly related to the general identification problem of dynamic factor models, where any of these procedures might find a local, and not a global, optimum. Generally, the more parameters, the harder it seems to be. Still, in our hardest case ($r = 2$ and $p = 2$) the $\beta_{i,k} = 0.2$ choice gives around and above 90 \% correct for $\omega = 0.2$ and around $80 \%$ for $\omega = 0.5$, which of course is lot higher than assuming full inclusion. We also see a general tendency that the method categorize better with more data, which is to be expected.

As a second evaluation criteria we look at estimation errors for factor loadings themselves. In order to compare estimations on the same scale we introduce scaling constants
\begin{align*}
    s_{i,j}^\star = s_{f_{j,t}}^\star/s_{\ti{y}_{i,t}}, \quad  \hat{s}_{i,j} = \hat{s}_{f_{j,t}}, \quad i=1, ..., n, j=1, ..., r, 
\end{align*}

where $s_{f_{j,t}}^\star$ and $\hat{s}_{f_{j,t}}$ are the standard deviations of the true factor $f_{j,t}$ and its estimated counterpart, respectively, and $s_{\ti{y}_{i,t}}$ is the standard deviation of variable $\ti{y}_{i,t}$. $s_{\ti{y}_{i,t}}$ is not included in $\hat{s}_{i,j}$ as data is already standardized in estimation. A measure of the root mean square error in loading estimation is given by the statistic
\begin{align*}
    E_\Lambda = \sqrt{\frac{1}{ns}\sum_{i=1}^n \sum_{j=1}^r \sum_{\ell = 0}^p \left(\lambda_{i,r\ell + j}^\star s_{i,j}^\star - \hat{\lambda}_{i,r\ell + j} \hat{s}_{i,j}\right)^2}, \numberthis \label{ELambda} 
\end{align*}

where $\lambda_{i,k}^\star$ is a true factor loading with its estimated counterpart $\hat{\lambda}_{i,k}$. For ML-estimation $\hat{\lambda}_{i,k}$ is the ML-estimate, and for VI-LS  $\hat{\lambda}_{i,k} = \E{q}{z_{i,k}\lambda_{i,k}} = b_{i,k}\mu_{\lambda_{i,k}}$. Table \ref{T:sim1LE} shows the average $E_\Lambda$ over simulations per specification scheme. 

\begin{table}[!ht]
\caption{Average factor loading RMSE using VI-LS and ML (Experiment 1)} 
\label{T:sim1LE}
\centering 
\renewcommand{\arraystretch}{1}
\resizebox{0.9\textwidth}{!}{
\begin{tabular}{l cc r ccc|c|r ccc|c|}
\cmidrule{4-13}													&	&	&	&	\multicolumn{4}{c}{$r=1, \;,p=0$}			&&	\multicolumn{4}{c}{$r=1,\;p=2$}		\\	\cline{5-8}	\cline{10-13}													$\omega$	&	$n$	&	$T$	&	$\beta:$	&	20\%	&	50\%	&	100 \%	&	ML	&&	20\%	&	50\%	&	100 \%	&	ML	\\	\hline		
\multirow{6}{*}{20\%}	&	50	&	100	&		&	\textbf{.048}	&	.065	&	.095	&	.096	&&	\textbf{.115}	&	.126	&	.145	&	.311	\\			
	&	50	&	200	&		&	\textbf{.034}	&	.046	&	.068	&	.068	&&	\textbf{.109}	&	.120	&	.134	&	.203	\\			
	&	100	&	100	&		&	\textbf{.048}	&	.065	&	.094	&	.096	&&	\textbf{.111}	&	.125	&	.148	&	.209	\\			
	&	100	&	200	&		&	\textbf{.032}	&	.046	&	.068	&	.068	&&	\textbf{.083}	&	.096	&	.116	&	.146	\\			
	&	400	&	100	&		&	\textbf{.047}	&	.064	&	.093	&	.096	&&	\textbf{.092}	&	.107	&	.131	&	.167	\\			
	&	400	&	200	&		&	\textbf{.032}	&	.045	&	.067	&	.067	&&	\textbf{.069}	&	.082	&	.100	&	.118	\\			
	&	&	&&	&	&		&	&	&&	&	&	\\														
\multirow{6}{*}{50\%}	&	50	&	100	&		&	\textbf{.063}	&	.070	&	.089	&	.089	&&	.141	&	\textbf{.140}	&	\textbf{.140}	&	.190	\\			
	&	50	&	200	&		&	\textbf{.041}	&	.048	&	.062	&	.062	&&	.129	&	.128	&	\textbf{.124}	&	.185	\\			
	&	100	&	100	&		&	\textbf{.062}	&	.069	&	.087	&	.088	&&	.119	&	\textbf{.116}	&	.124	&	.151	\\			
	&	100	&	200	&		&	\textbf{.041}	&	.048	&	.062	&	.062	&&	\textbf{.089}	&	.091	&	.096	&	.118	\\			
	&	400	&	100	&		&	\textbf{.062}	&	.069	&	.087	&	.088	&&	.106	&	\textbf{.105}	&	.119	&	.137	\\			
	&	400	&	200	&		&	\textbf{.040}	&	.047	&	.061	&	.061	&&	\textbf{.078}	&	.080	&	.089	&	.096	\\			
	&	&	&&	&	&		&	&&	&	&	&	\\														
\multirow{6}{*}{100\%}	&	50	&	100	&		&	.077	&	.073	&	\textbf{.072}	&	\textbf{.072}	&&	.148	&	.130	&	\textbf{.120}	&	.138	\\			
	&	50	&	200	&		&	\textbf{.051}	&	\textbf{.051}	&	\textbf{.051}	&	\textbf{.051}	&&	.102	&	.091	&	\textbf{.083}	&	.093	\\			
	&	100	&	100	&		&	.078	&	.075	&	\textbf{.073}	&	\textbf{.073}	&&	.143	&	.124	&	\textbf{.114}	&	.122	\\			
	&	100	&	200	&		&	.051	&	\textbf{.050}	&	\textbf{.050}	&	\textbf{.050}	&&	.094	&	.084	&	\textbf{.078}	&	.080	\\			
	&	400	&	100	&		&	.077	&	.074	&	\textbf{.072}	&	\textbf{.072}	&&	.122	&	.105	&	\textbf{.098}	&	.102	\\			
	&	400	&	200	&		&	\textbf{.051}	&	\textbf{.051}	&	\textbf{.051}	&	\textbf{.051}	&&	.085	&	.075	&	\textbf{.070}	&	.085	\\ \hline			
	&	&	&	&	\multicolumn{4}{c}{$r=2,\;,p=0$}			&&	\multicolumn{4}{c}{$r=2,\;p=2$}		\\	\cline{5-8}	\cline{10-13}													
$\omega$	&	$n$	&	$T$	&	$\beta:$	&	20\%	&	50\%	&	100 \%	&	ML	&&	20\%	&	50\%	&	100	\%	&	ML	\\	\hline	
\multirow{6}{*}{20\%}	&	50	&	100	&		&	\textbf{.053}	&	.075	&	.121	&	.134	&&	\textbf{.112}	&	.119	&	.153	&	.447	\\			
	&	50	&	200	&		&	\textbf{.042}	&	.057	&	.090	&	.101	&&	\textbf{.089}	&	.104	&	.130	&	.272	\\			
	&	100	&	100	&		&	\textbf{.054}	&	.074	&	.130	&	.142	&&	\textbf{.091}	&	.101	&	.139	&	.403	\\			
	&	100	&	200	&		&	\textbf{.041}	&	.062	&	.095	&	.108	&&	\textbf{.086}	&	.098	&	.127	&	.287	\\			
	&	400	&	100	&		&	\textbf{.049}	&	.076	&	.138	&	.150	&&	\textbf{.071}	&	.085	&	.135	&	.262	\\			
	&	400	&	200	&		&	\textbf{.042}	&	.068	&	.109	&	.122	&&	\textbf{.058}	&	.072	&	.106	&	.129	\\			
	&	&	&&	&	&	&	&&	&	&	&	\\															
\multirow{6}{*}{50\%}	&	50	&	100	&		&	\textbf{.095}	&	.108	&	.148	&	.180	&&	.126	&	\textbf{.120}	&	.146	&	.314	\\			
	&	50	&	200	&		&	\textbf{.079}	&	.089	&	.130	&	.144	&&	\textbf{.100}	&	.101	&	.127	&	.232	\\			
	&	100	&	100	&		&	\textbf{.091}	&	.105	&	.152	&	.180	&&	.100	&	\textbf{.098}	&	.137	&	.395	\\			
	&	100	&	200	&		&	\textbf{.081}	&	.096	&	.137	&	.160	&&	\textbf{.084}	&	.089	&	.120	&	.195	\\			
	&	400	&	100	&		&	\textbf{.089}	&	.114	&	.184	&	.199	&&	\textbf{.093}	&	.095	&	.135	&	.266	\\			
	&	400	&	200	&		&	\textbf{.094}	&	.114	&	.153	&	.166	&&	\textbf{.072}	&	.078	&	.110	&	.144	\\			
&	&	&&	&	&	&	&&	&	&	&	\\																
\multirow{6}{*}{100\%}	&	50	&	100	&		&	.193	&	.188	&	\textbf{.170}	&	.181	&&	.169	&	.155	&	\textbf{.149}	&	.286	\\			
	&	50	&	200	&		&	.171	&	.170	&	\textbf{.161}	&	.169	&&	.144	&	.136	&	\textbf{.129}	&	.166	\\			
	&	100	&	100	&		&	.195	&	.187	&	\textbf{.179}	&	.188	&&	.160	&	.143	&	\textbf{.137}	&	.173	\\			
	&	100	&	200	&		&	.165	&	.163	&	\textbf{.158}	&	.163	&&	.137	&	.126	&	\textbf{.119}	&	.139	\\			
	&	400	&	100	&		&	.166	&	.162	&	\textbf{.154}	&	.159	&&	.152	&	.135	&	\textbf{.131}	&	.193	\\			
	&	400	&	200	&		&	.151	&	.147	&	\textbf{.142}	&	.146	&&	.119	&	.109	&	\textbf{.103}	&	.127	\\ \hline			
\end{tabular}}
    \begin{minipage}{\textwidth}
    \small 
    \emph{Note: Bold-face denotes first place (or shared first place) in evaluation criteria. $n$ is number of variables, $T$ is sample size, $r$ is number of factors, $p$ is number of factor lags, $\omega$ is share of included factor loadings in simulation, $\beta$ is prior probability of inclusion per factor loading. }
    \end{minipage}
\end{table}

In case of higher sparsity, VI-LS with small inclusion prior show much smaller errors. VI-LS perform better than ML; when we have sparsity the benefits are substantial, and holds over different sample sizes, number of variables, factors and lag-lengths.

Interestingly, the opposite is less true. To put more clearly, the costs of assuming sparsity when there is no sparsity leads to somewhat increased errors, but the difference is not as big. Also, these differences seemingly go away with more data. Regardless, VI is still preferable to ML. 

If a researcher is highly unsure about the degree of sparsity, and unluckily has to resort to one single model specification, these result suggests that assuming sparsity a-priori is the least risky move.  

The third evaluation looks at factor estimation precision. In contrast to the previous two evaluations, this is not impeded by issues of identifying particular rotations, as long as the factor space is correctly estimated.

Define true factors $f_t^\star$ and estimated factors $\hat{f}_t$ with corresponding stacked matrices: 
\begin{align*}
    F^\star = \begin{bmatrix} {f_1^\star}' \\
    {f_2^\star}' \\ \vdots \\ {f_T^\star}' \end{bmatrix}\text{ and } \hat{F} = \begin{bmatrix} \hat{f}_1' \\
    \hat{f}_2' \\ \vdots \\ \hat{f}_T' \end{bmatrix}. 
\end{align*}

We follow \cite{DozEtAl2012} \citep[see also][]{StockWatson2002a,BanburaetAl2014} and using the trace of the \textit{coefficient of determination} (typically denoted $R^2$), given from a multivariate regression of $\hat{F}$ onto $F^\star$. This factor precision statistic is given by
\begin{align*}
     P_F = \frac{\Tr\big({F^\star}'\hat{P}{F^\star}  \big)}{\Tr\big({F^\star}'F^\star\big)}, \numberthis \label{precisonstatistic} 
\end{align*}
where we have the projection matrix $\hat{P} = \hat{F}'\big(\hat{F}'\hat{F}\big)^{-1}\hat{F}$.

Table \ref{T:sim1PS} shows the average factor precision \eqref{precisonstatistic} over simulations for each specification scheme.

\begin{table}[!ht]
\caption{Average factor estimation precision statistics using VI-LS and ML, in percent (Experiment 1)} 
\label{T:sim1PS}
\centering
\renewcommand{\arraystretch}{1}
\resizebox{0.9\textwidth}{!}{
\begin{tabular}{l cc r ccc|c|r ccc|c|}
\cmidrule{4-13}										
&	&	&	&	\multicolumn{4}{c}{$r=1, \;p=0$}			&&	\multicolumn{4}{c}{$r=1,\;p=2$}		\\	\cline{5-8}	\cline{10-13}													
$\omega$	&	$n$	&	$T$	&	$\beta:$	&	20\%	&	50\%	&	100 \%	&	ML	&&	20\%	&	50\%	&	100 \%	&	ML	\\	\hline	
\multirow{6}{*}{20\%}	&	50	&	100	&		&	\textbf{90.65}	&	90.58	&	90.38	&	90.47	&&	81.33	&	82.28	&	\textbf{85.15}	&	79.74	\\		
	&	50	&	200	&		&	\textbf{92.99}	&	92.96	&	92.88	&	92.92	&&	80.01	&	80.42	&	\textbf{81.77}	&	79.28	\\		
	&	100	&	100	&		&	\textbf{94.94}	&	94.91	&	94.84	&	94.88	&&	81.89	&	82.35	&	\textbf{83.84}	&	81.69	\\		
	&	100	&	200	&		&	\textbf{95.41}	&	95.39	&	95.36	&	95.37	&&	87.19	&	87.20	&	\textbf{88.32}	&	86.59	\\		
	&	400	&	100	&		&	\textbf{96.94}	&	96.93	&	96.91	&	96.92	&&	87.36	&	87.39	&	\textbf{89.39}	&	87.51	\\		
	&	400	&	200	&		&	\textbf{98.00}	&	97.99	&	97.99	&	97.99	&&	90.04	&	90.10	&	\textbf{91.61}	&	90.06	\\		
	&	&	&&	&	&		&	&	&&	&	&	\\													
\multirow{6}{*}{50\%}	&	50	&	100	&		&	\textbf{95.79}	&	\textbf{95.79}	&	\textbf{95.79}	&	\textbf{95.79}	&&	84.32	&	84.38	&	\textbf{86.37}	&	83.72	\\		
	&	50	&	200	&		&	\textbf{96.34}	&	\textbf{96.34}	&	\textbf{96.34}	&	\textbf{96.34}	&&	83.61	&	83.97	&	\textbf{86.24}	&	83.29	\\		
	&	100	&	100	&		&	\textbf{96.38}	&	\textbf{96.38}	&	96.37	&	96.37	&&	91.14	&	91.59	&	\textbf{92.58}	&	91.10	\\		
	&	100	&	200	&		&	\textbf{97.74}	&	\textbf{97.74}	&	\textbf{97.74}	&	\textbf{97.74}	&&	92.42	&	92.49	&	\textbf{93.38}	&	92.29	\\		
	&	400	&	100	&		&	\textbf{97.38}	&	\textbf{97.38}	&	\textbf{97.38}	&	\textbf{97.38}	&&	92.51	&	92.51	&	\textbf{92.62}	&	92.57	\\		
	&	400	&	200	&		&	\textbf{98.19}	&	\textbf{98.19}	&	98.18	&	\textbf{98.19}	&&	94.75	&	94.75	&	\textbf{95.25}	&	94.75	\\		
	&	&	&&	&	&		&	&&	&	&	&	\\													
\multirow{6}{*}{100\%}	&	50	&	100	&		&	\textbf{95.93}	&	\textbf{95.93}	&	\textbf{95.93}	&	\textbf{95.93}	&&	92.50	&	92.59	&	\textbf{92.70}	&	92.14	\\		
	&	50	&	200	&		&	\textbf{97.03}	&	\textbf{97.03}	&	\textbf{97.03}	&	\textbf{97.03}	&&	95.04	&	95.09	&	\textbf{95.43}	&	95.03	\\		
	&	100	&	100	&		&	\textbf{96.62}	&	\textbf{96.62}	&	\textbf{96.62}	&	\textbf{96.62}	&&	93.35	&	93.38	&	\textbf{93.47}	&	93.43	\\		
	&	100	&	200	&		&	\textbf{97.97}	&	\textbf{97.97}	&	\textbf{97.97}	&	\textbf{97.97}	&&	97.42	&	97.43	&	\textbf{97.45}	&	97.45	\\		
	&	400	&	100	&		&	\textbf{97.13}	&	\textbf{97.13}	&	\textbf{97.13}	&	\textbf{97.13}	&&	\textbf{96.81}	&	\textbf{96.81}	&	96.48	&	96.37	\\		
	&	400	&	200	&		&	\textbf{98.67}	&	\textbf{98.67}	&	\textbf{98.67}	&	\textbf{98.67}	&&	98.31	&	98.32	&	\textbf{98.33}	&	98.32	\\ \hline		
	&	&	&	&	\multicolumn{4}{c}{$r=2,\;p=0$}			&&	\multicolumn{4}{c}{$r=2,\;p=2$}		\\	\cline{5-8}	\cline{10-13}												
$\omega$	&	$n$	&	$T$	&	$\beta:$	&	20\%	&	50\%	&	100 \%	&	ML	&&	20\%	&	50\%	&	100	\%	&	ML	\\	\hline
\multirow{6}{*}{20\%}	&	50	&	100	&		&	\textbf{90.28}	&	90.15	&	89.77	&	90.03	&&	84.25	&	85.35	&	\textbf{88.13}	&	78.70	\\		
	&	50	&	200	&		&	\textbf{91.58}	&	91.52	&	91.38	&	91.46	&&	85.82	&	85.76	&	\textbf{86.67}	&	83.30	\\		
	&	100	&	100	&		&	\textbf{94.75}	&	94.72	&	94.59	&	94.66	&&	89.65	&	89.94	&	\textbf{91.09}	&	85.62	\\		
	&	100	&	200	&		&	\textbf{94.95}	&	94.93	&	94.88	&	94.90	&&	88.06	&	88.51	&	\textbf{89.51}	&	85.29	\\		
	&	400	&	100	&		&	\textbf{96.23}	&	96.22	&	96.20	&	96.22	&&	93.91	&	93.97	&	\textbf{94.07}	&	92.28	\\		
	&	400	&	200	&		&	\textbf{97.46}	&	\textbf{97.46}	&	97.45	&	\textbf{97.46}	&&	95.52	&	95.52	&	\textbf{95.96}	&	95.19	\\		
	&	&	&&	&	&	&	&&	&	&	&	\\														
\multirow{6}{*}{50\%}	&	50	&	100	&		&	\textbf{94.14}	&	94.13	&	94.11	&	94.12	&&	88.90	&	89.67	&	\textbf{90.85}	&	86.54	\\		
	&	50	&	200	&		&	\textbf{95.37}	&	\textbf{95.37}	&	95.36	&	95.36	&&	91.54	&	92.36	&	\textbf{92.96}	&	90.73	\\		
	&	100	&	100	&		&	\textbf{95.62}	&	95.61	&	95.60	&	95.61	&&	94.33	&	94.29	&	\textbf{94.67}	&	93.29	\\		
	&	100	&	200	&		&	\textbf{96.98}	&	\textbf{96.98}	&	96.97	&	96.97	&&	94.53	&	94.64	&	\textbf{94.87}	&	93.34	\\		
	&	400	&	100	&		&	\textbf{97.20}	&	\textbf{97.20}	&	\textbf{97.20}	&	\textbf{97.20}	&&	96.07	&	96.00	&	\textbf{96.34}	&	95.42	\\		
	&	400	&	200	&		&	\textbf{98.49}	&	\textbf{98.49}	&	\textbf{98.49}	&	\textbf{98.49}	&&	97.48	&	97.46	&	\textbf{97.52}	&	96.38	\\		
&	&	&&	&	&	&	&&	&	&	&	\\															
\multirow{6}{*}{100\%}	&	50	&	100	&		&	94.33	&	94.34	&	\textbf{94.35}	&	\textbf{94.35}	&&	91.87	&	92.12	&	\textbf{92.65}	&	91.41	\\		
	&	50	&	200	&		&	96.19	&	\textbf{96.20}	&	\textbf{96.20}	&	\textbf{96.20}	&&	93.58	&	93.69	&	\textbf{93.90}	&	93.05	\\		
	&	100	&	100	&		&	95.34	&	\textbf{95.35}	&	\textbf{95.35}	&	\textbf{95.35}	&&	94.90	&	95.32	&	\textbf{95.78}	&	94.95	\\		
	&	100	&	200	&		&	\textbf{97.70}	&	\textbf{97.70}	&	\textbf{97.70}	&	\textbf{97.70}	&&	95.23	&	95.30	&	\textbf{95.86}	&	95.12	\\		
	&	400	&	100	&		&	\textbf{96.77}	&	\textbf{96.77}	&	\textbf{96.77}	&	\textbf{96.77}	&&	96.55	&	96.58	&	\textbf{96.71}	&	96.24	\\		
	&	400	&	200	&		&	\textbf{98.37}	&	\textbf{98.37}	&	\textbf{98.37}	&	\textbf{98.37}	&&	97.65	&	97.67	&	\textbf{97.91}	&	97.67	\\ \hline		
\end{tabular}}
    \begin{minipage}{\textwidth}
    \small 
    \emph{Note: Bold-face denotes first place (or shared first place) in evaluation criteria. $n$ is number of variables, $T$ is sample size, $r$ is number of factors, $p$ is number of factor lags, $\omega$ is share of included factor loadings in simulation, $\beta$ is prior probability of inclusion per factor loading. }
    \end{minipage}
\end{table}

The estimations yield often quite similar precisions, where VI-LS with $\beta_{i,k} = 0.2$ has a (small) advantage in the no-lag cases with $\omega=0.2$. In the no-lag case with more inclusion, the precisions are almost indistinguishable. The largest differences are seen in the lag-cases with partial inclusion, where ML often perform worst and VI with full inclusion often perform best. 

 This study infer no particular benefit of ML-estimation, even with full loading inclusion. Added to that is the fact that VI provides uncertainty estimates of parameters; the usage of ML becomes even less favoured. VI is clearly preferable and by VI-LS we obtain additional flexibility.
 
 In some of the simulations, assuming sparsity cost us some factor precision. This is sometimes due to the procedure converging to local, sub-optimal, optimum, which in our case reduce the average precision of VI-LS somewhat. However, it is highly reasonable to think that we can improve upon these results with a more elaborated prior selection scheme, carefully tailored to each individual simulation, perhaps using cross-validation under a training sample or something similar. Spending more time for any individual estimations, the researcher could investigate more of the optimization space, trying different start-values and perhaps remedy the situation of lesser local optimums. In this experiment we have not followed such tailor-made rules, but act rather heuristically. 
 
However, this experiment shows is that even heuristic priors and start values can provide reasonable categorization and sparse solutions. Also, the estimates can be obtained quickly. Estimation for the smallest case ($n=50$, $T=100$, $r=1$ and $p=0$) took on average less than 0.9 seconds and the largest case ($n=400$, $T=200$, $r=2$ and $p=2$) took on average about 31.7 seconds, on a standard laptop.\footnote{Intel core i7-10870H CPU @ 2.20GHz with 32GB RAM} This includes re-running according to Algorithm \ref{algoRerunVIDFM}. What is more: the computational time can likely be reduced further by less strict tolerance levels, with almost equivalent results. Additionally, it is typically much quicker updating an already existing model with new data by using previous estimates as starting values; our figures does not take this aspect into account. 

\subsection{Simulation experiment 2}

In the second experiment, we investigate investigate even larger and sparser models. We simulate two DFMs using \eqref{sim1}-\eqref{simlast}, with $n=800$, $T=250$, $r=4$ and $p=2$. Thus, we have $n  \times s = 9600$ factor loadings, each of which may or may not be included. In both cases we use prior $\beta = \frac{1}{10} \mathbf{1}_{n \times s}$, with the remaining chosen as in Section \ref{sec.sim1}. The simulations differ in degree of sparsity, where in the first $\omega=0.1$ and second $\omega=0.025$, i.e, the second simulation is sparser than our prior belief. We also impose a missing data pattern, according to:
\begin{itemize}
    \item First 200 variables are fully available. 
    \item Second 200 variables are of "lower frequency", i.e. only each third observation is available.
    \item Third 200 variables start at a random point, where we uniformly choose a variable, take away the first current available observation, and iterate until 20 \% of the observations are missing.  
    \item Fourth 200 variables have random missing pattern, 20 \% chosen uniformly over all observations
\end{itemize}

\begin{figure}[!ht]
    \centering
     \caption{Data availability pattern}
    \includegraphics[width=0.70\textwidth]{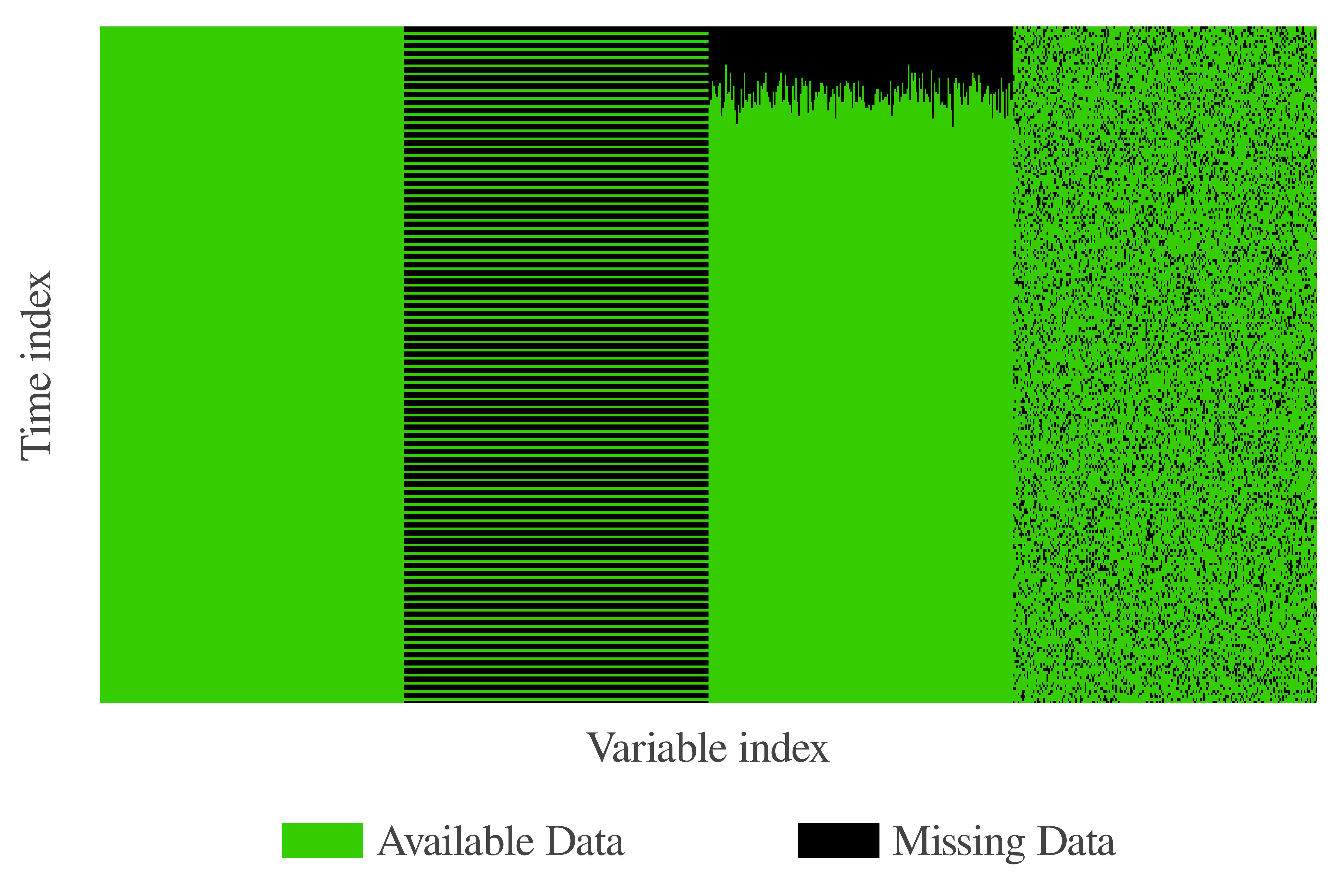}
    \label{mdatafig}
\end{figure}

Figure \ref{mdatafig} shows the missing pattern for the first simulation (the second has a very similar pattern). Data availability is denoted by green, and missing data by black. The column-dimension shows variable index, ordered according to the four blocks described above. The row-dimension is time index.

Table \ref{T:sim2} shows factor loading RMSE \eqref{PZ} and factor precision \eqref{precisonstatistic} for Simulation experiment 2. 

\begin{table}[!ht]
\centering
 \begin{threeparttable}
\caption{Loading errors and factor space precision}
\label{T:sim2}
\begin{tabular}{l|cc|cc|}
&	\multicolumn{2}{|c|}{Factor loading RMSE}			&	\multicolumn{2}{|c|}{Factor space precision}			\\	\hline
$\omega$	&	VI-LS	&	ML	&	VI-LS	&	ML	\\	\hline
10 \%	&	.036	&	.139	&	.993	&	.991	\\	
2.5 \%	&	.024	&	.111	&	.982	&	.971	\\	\hline
\end{tabular}
\begin{tablenotes}
      \small
      \item Remark: In both $\omega$-cases, VI-LS has loading inclusion probability 10 \%
    \end{tablenotes}
    \end{threeparttable}
\end{table}
VI-LS yields superior results, especially in terms of loading estimation. The RMSE of factor loadings are much higher for ML.

\begin{figure}[H]
    \centering
    \includegraphics[width=1\textwidth]{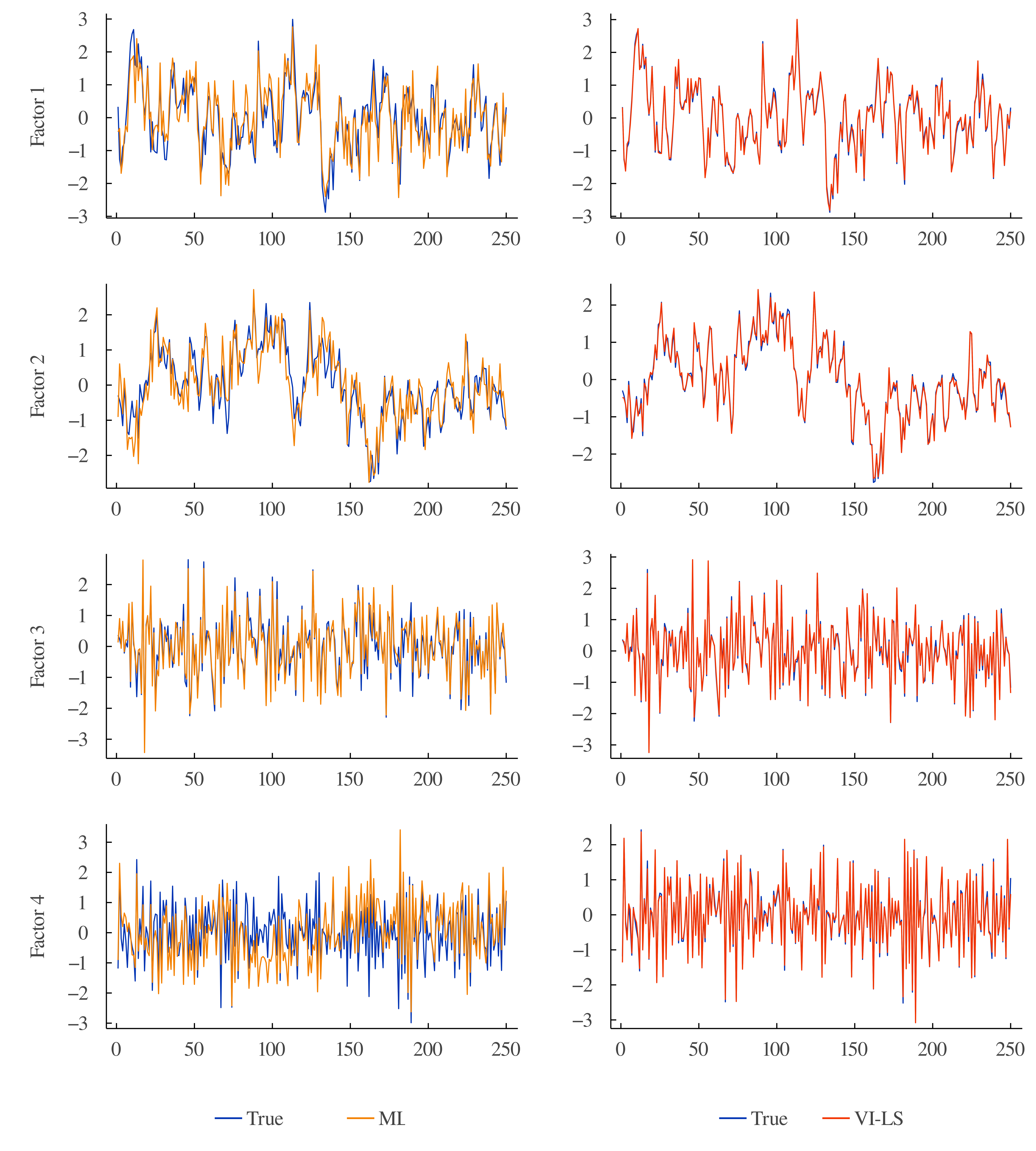}
         \caption{True factors with 10 \% loading inclusion and estimated factors, VI-LS with 10 \% prior loading inclusion probability (Experiment 2)}
    \label{F1}
\end{figure}

\begin{figure}[H]
    \centering
    \includegraphics[width=1\textwidth]{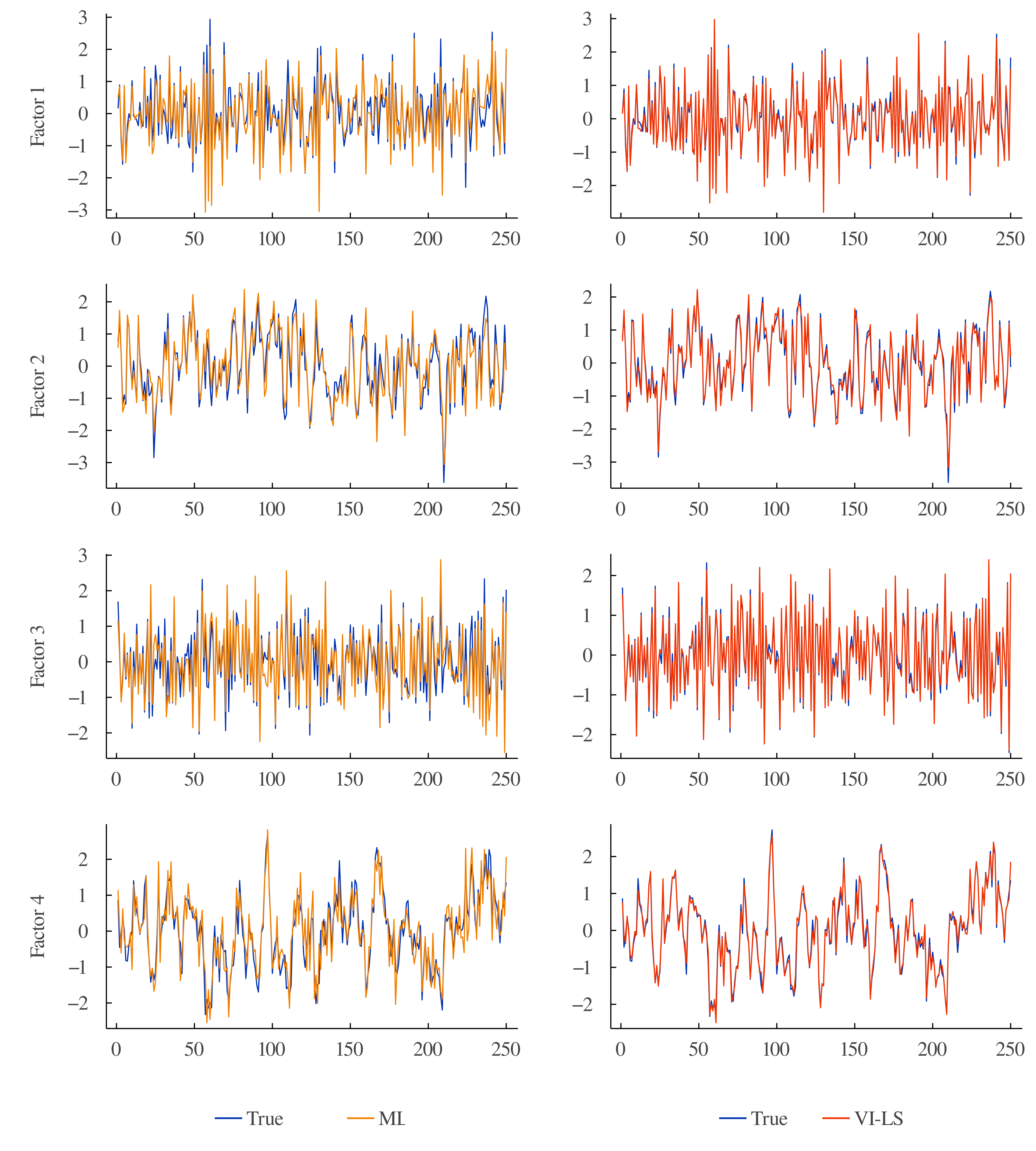}
         \caption{True factors with 2.5 \% loading inclusion and estimated factors, VI-LS 10 \% prior loading inclusion probability (Experiment 2)}
    \label{F2}
\end{figure}

VI-LS also provide somewhat better factor space precision, but it does not differ much from ML. Although when studying the individual factors (Figure \ref{F1} and \ref{F2}) we see a much clearer advantage for VI-LS. VI-LS (right plots) capture the factor patterns better than ML (left plots). Putting together the facts that (i) loading estimation is much better for VI-LS, (ii) factor patterns are much clearly captured and (iii), factor space estimation do not differ that much, we can deduce the VI-LS benefit: factor identification.

\begin{figure}[H]
    \centering
    \includegraphics[width=1\textwidth]{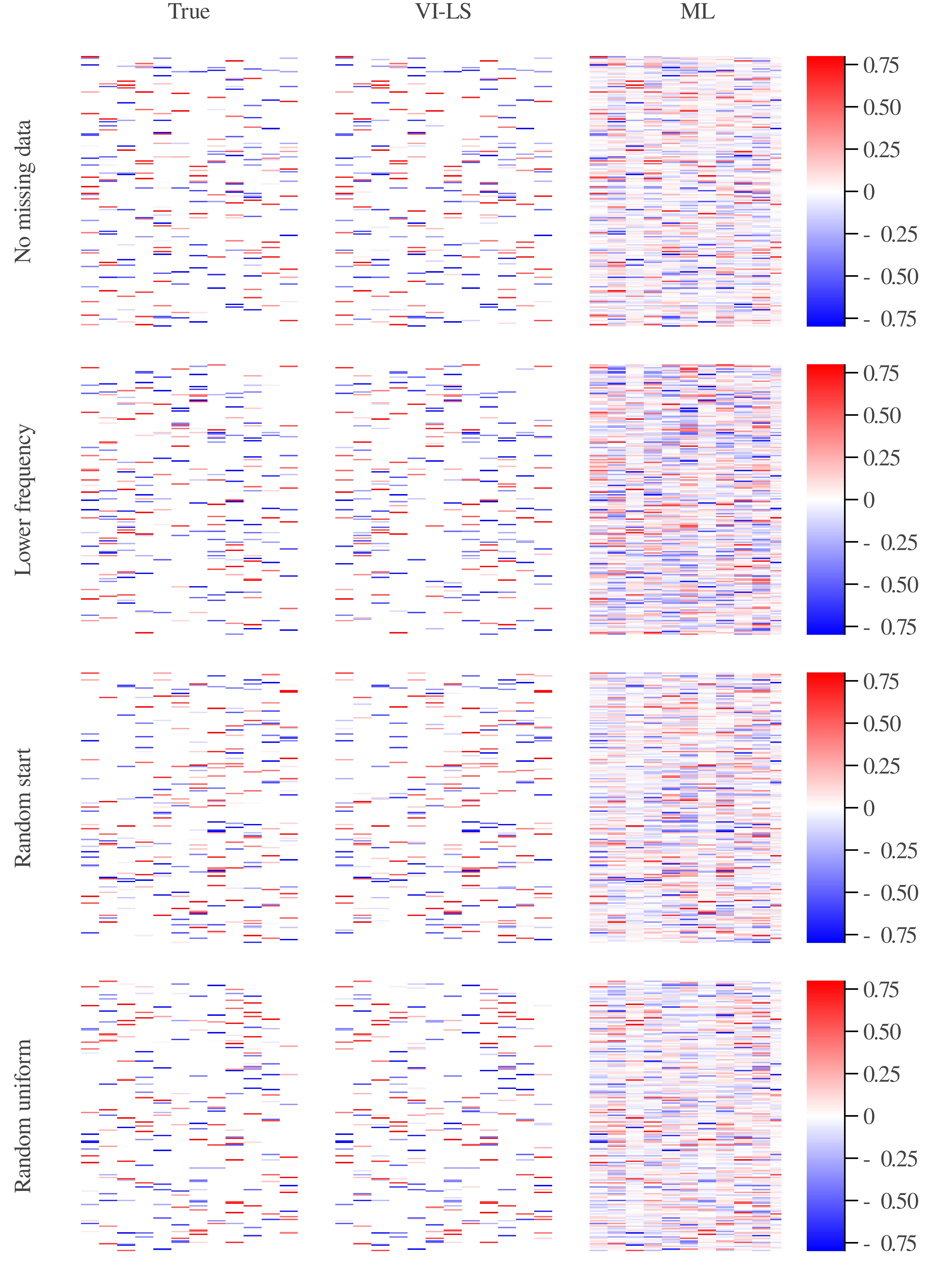}
         \caption{True factor loading matrices with 10 \% loading inclusion, VI-Ls with  10 \% prior loading inclusion probability (Experiment 2)}
    \label{FL1}
\end{figure}

\begin{figure}[H]
    \centering
    \includegraphics[width=1\textwidth]{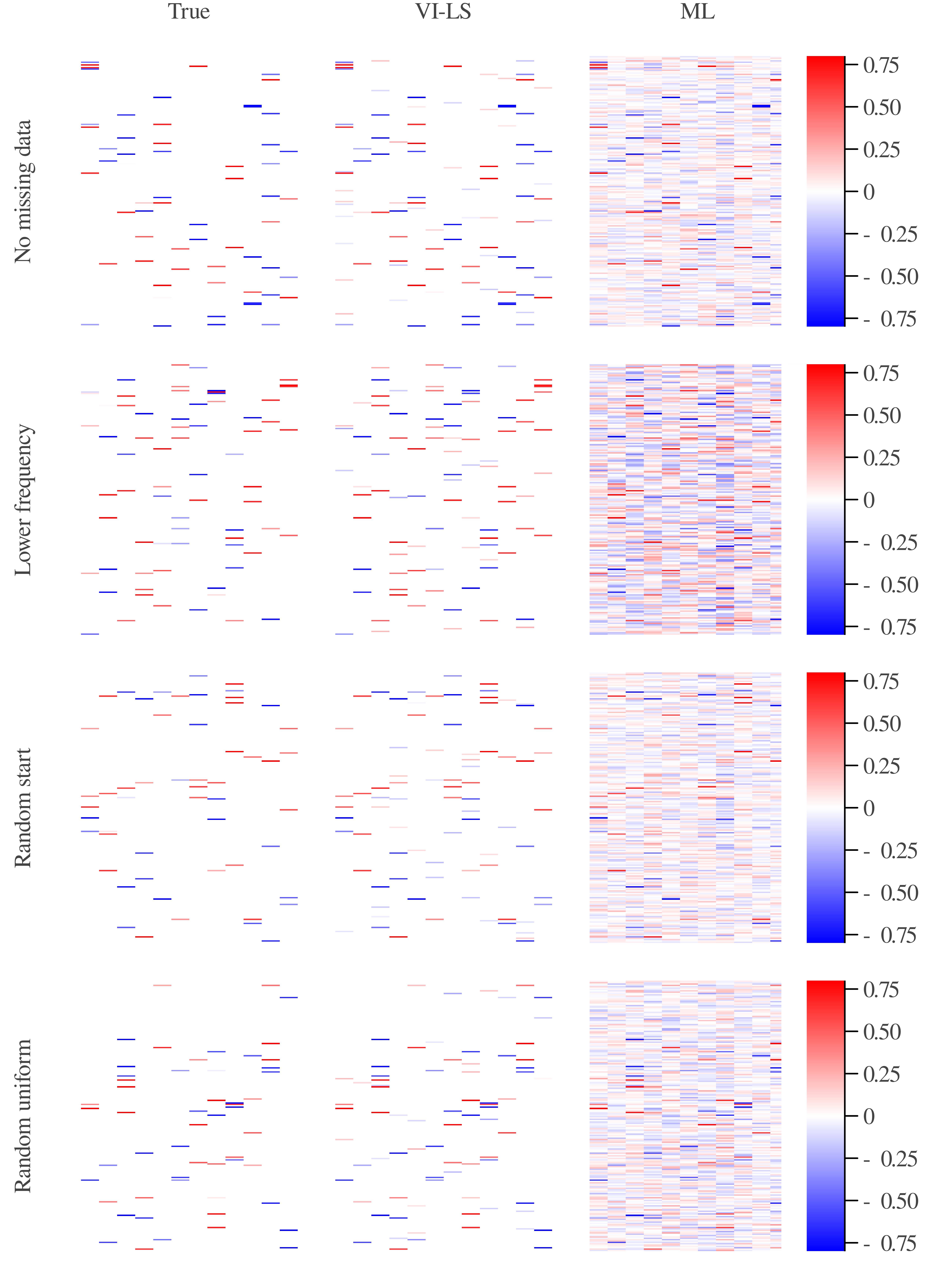}
             \caption{True factor loading matrices with 2.5 \% loading inclusion, VI-LS with 10 \% prior loading inclusion probability (Experiment 2)}
    \label{FL2}
\end{figure}

This becomes even more clear when studying the factor loadings sparsity pattern in Figures \ref{FL1} and \ref{FL2}.   VI-LS shows sparsity, very much inline with the true pattern. The results holds over all studied availability patterns. 

When true sparsity is higher than the prior belief (Figure \ref{FL2}) we see some feint false estimation inclusions. In such cases however, loadings are typically estimated to be quite small. 

ML loading estimates on the other hand are a lot more cluttered. They do not find the correct sparse solution, again highlighting the superiority of VI-LS in factor identification.   
\newpage

\newpage 
\section{Conclusions}

We have developed a quick algorithm to estimate sparse DFMs, providing posterior approximations and applicable to data sets with missing data patterns, without resorting to simulation techniques. Specifically, we deal with sparsity by estimating loading indicators with a spike-and-slab prior scheme, and using a mean field algorithm to provide analytical solutions in recursive steps, similar to Expectation Maximization. 

Monte Carlo simulation experiments show some clear benefits of our approach in case of sparsity.VI is preferable to ML generally, and the spike-and-slab scheme gives an added flexibility in model construction to deal with sparsity situations. The most obvious benefit is the enhanced ability to identify the correct system. 

As the estimates can be provided quickly, the algorithm is a novel tool effectively dealing with sparsity for practical forecasting with limited time and computational constraints. It reduces the gap between fast estimations and deeper analytics. 

Future research could study different prior schemes, where we believe there may be a lot to unpack and improve. 

Additionally, loading inclusion priors can be used to introduce prior structural beliefs, where a researcher could direct the data to certain structures in a pragmatic way. For example, a macro-economist could use their preconceived notions of structural shocks derived from macroeconomic theory, but impose them only to a chosen degree of certainty, directing the interpretation of factors to certain features. This particular aspect should be worth investigating for future research, as it may provide a suitable compromise between purely structural models and more empirical models.

\bibliography{references}

\newpage 
\begin{appendix} 
\section{Variational density derivations} \label{sec:A}
Adding loading inclusion and its priors to the joint density of S21, and more flexible priors for $\Lambda$ and $\Phi$, we get the log joint density
\begin{align*}
    \ln p\left(\Omega, F, \theta, Z\right) &= C - \frac{1}{2}\sum_{i=1}^n T_i \ln \sigma_{\epsilon_i}^2 - \sum_{i=1}^n\sum_{t=1}^T \frac{a_{i,t}}{2\sigma_i^2}\left(\ti{y}_{i,t} - (z \circ \lambda_i)'F_t\right)^2 \\
    &- \frac{T}{2}\sum_{j=1}^r \ln \sigma_{\ti{u}_j}^2 - \frac{1}{2}F_0' V_{F_0}^{-1} F_0 - \frac{1}{2}\sum_{t=1}^T\left(F_t - \widetilde{\Phi}F_{t-1}\right)'\widetilde{\Sigma}_\ti{u}^{-1}\left(F_t - \widetilde{\Phi}F_{t-1}\right) \\
    &-\frac{s}{2}\sum_{i=1}^n \ln \sigma_{\epsilon_i}^2  -\frac{1}{2} \sum_{i=1}^n \frac{1}{\sigma^2_{\epsilon_i}} \lambda_i'V_{\lambda_i}^{-1}\lambda_i - \sum_{i=1}^n \left(1 + \nu_{\epsilon_i}/2\right) \ln \sigma^2_{\epsilon_i} - \sum_{i=1}^n \frac{\nu_{\epsilon_i} \tau^2_{\epsilon_i}}{2\sigma^2_{\epsilon_i}}  \\ 
    &-\frac{s}{2}\sum_{j=1}^r \ln \sigma_{\ti{u}_j}^2  -\frac{1}{2} \sum_{j=1}^r \frac{1}{\sigma^2_{\ti{u}_j}} \phi_i'V_{\phi_i}^{-1}\phi_i - \sum_{j=1}^r \left(1 + \nu_{\ti{u}_j}/2\right) \ln \sigma^2_{\ti{u}_j} - \sum_{j=1}^r \frac{\nu_{\ti{u}_j} \tau^2_{\ti{u}_j}}{2\sigma^2_{\ti{u}_j}}  \\ 
    &+ \sum_{i=1}^n\sum_{k=1}^s z_{i.k}\text{logit}(\beta_{i,k})\numberthis \label{jointdens}
\end{align*}
where $C$ is a constant. 

\subsection{Density of $\{\Lambda, \Sigma_\epsilon\}$} \label{sec:A1}
This section is similar to the derivations of S21.\footnote{See Appendix A.2 therein} By independence following the definition of exact factor model we know that $q(\Lambda, \Sigma_\epsilon) = \prod_{i=1}^n q(\lambda_i,\sigma^2_{\epsilon_i}) $. Log variational density of $\{\lambda_i, \sigma^2_i\}$ is given by 
\begin{align*}
    \ln q\left(\lambda_i, \sigma^2_i \right) &= \E{q(F,Z)}{\ln p(Y_A, F, \theta, Z)} \\
    &=  C_{\lambda_i,\sigma^2_{\epsilon_i}} - \frac{1}{2} T_i \ln \sigma_{\epsilon_i}^2 - \sum_{t=1}^T  \frac{a_{i,t}}{2\sigma_{\epsilon_i}^2}\E{q\left(F,Z\right)}{\left(\ti{y}_{i,t} - \lambda_i'(z_i \circ F_t)\right)^2} \\
   &\quad -\frac{s}{2}\ln \sigma_{\epsilon_i}^2  - \frac{1}{2\sigma^2_{\epsilon_i}} \lambda_i'V_{\lambda_i}^{-1}\lambda_i - \left(1 + \nu_{\epsilon_i}/2\right) \ln \sigma^2_{\epsilon_i} -  \frac{\nu_{\epsilon_i} \tau^2_{\epsilon_i}}{2\sigma^2_{\epsilon_i}}, 
\end{align*} 
and where $C_{\lambda_i,\sigma^2_{\epsilon_i}}$ is constant in terms of $\lambda_i$ and $\sigma^2_{\epsilon_i}$. 

The expectation-term is given by
\begin{align*}
    &\sum_{t=1}^T  \frac{a_{i,t}}{2\sigma_{\epsilon_i}^2}\E{q\left(F\right)}{\left(\ti{y}_{i,t} - \lambda_i'(z_i \circ F_t)\right)^2} \\ 
    &\qquad = \frac{1}{2\sigma^2_{\epsilon_i}} \left(\sum_{t=1}^Ta_{i,t}\ti{y}_{i,t}^2 - 2\lambda_i' \sum_{t=1}^T\E{q(F,Z)}{z_i \circ F_t}a_{i,t}\ti{y}_{i,t} + \lambda_i'\sum_{t=1}^Ta_{i,t}\E{q(F,Z)}{(z_i \circ F_t)(z_i \circ F_t)'}, \lambda_i \right) \numberthis \label{stepA1}
\end{align*}
where 
\begin{align*}
    \E{q(F,Z)}{z_i \circ F_t} &= \E{q(Z)}{z_i} \circ \E{q(F)}{F_t} = b_i \circ \E{q(F)}{F_t}, \\ 
    \E{q(F,Z)}{(z_i \circ F_t)(z_i \circ F_t)'} &= \E{q(Z)}{z_i z_i'} \circ \E{q(F)}{F_t F_t'} =  P_i \circ \E{q(F)}{F_t F_t'}.
\end{align*}
Inserting \eqref{stepA1} gives
\begin{align*}
    \ln q(\lambda_i, \sigma^2_{\epsilon_i})&=   C_{\lambda_i,\sigma^2_{\epsilon_i}} - \frac{1}{2} T_i \ln \sigma_{\epsilon_i}^2 - \frac{s}{2}\ln \sigma_{\epsilon_i}^2  - \frac{1}{2\sigma^2_{\epsilon_i}} \lambda_i'V^{-1}\lambda_i - \left(1 + \nu_{\epsilon_i}/2\right) \ln \sigma^2_{\epsilon_i} - \frac{\nu_{\epsilon_i} \tau^2_{\epsilon_i}}{2\sigma^2_{\epsilon_i}} \\
   &\quad -\frac{1}{2\sigma^2_{\epsilon_i}} \left(\sum_{t=1}^T a_{i,t}\ti{y}_{i,t}^2 - 2\lambda_i'\left(b_i \circ \sum_{t=1}^T \E{q(F)}{F_t} a_{i,t} \ti{y}_{i,t}\right) + \lambda_i'\left(P_i \circ \sum_{t_1}^Ta_{i,t}\E{q(F)}{F_t F_t'}\right)\lambda_i \right) ,
\end{align*}
Introducing objects
\begin{align*}
    g_i = \sum_{t=1}^T \E{q(F)}{F_t} a_{i,t}  \ti{y}_{i,t} \quad \text{and} \quad  Q_i = \sum_{t=1}^T a_{i,t}\E{q(F)}{F_t F_t'} 
\end{align*}
and rearranging terms (e.g. gather all $\lambda_i$-terms) yields
\begin{align*}
    \ln q(\lambda_i, \sigma^2_{\epsilon_i}) &=   C_{\lambda_i,\sigma^2_i} -  \frac{1}{2\sigma_i^2}\Bigg(-2  \lambda_i'\left(b_i \circ g_i\right) + \lambda_i'\left(P_i \circ Q_i + V_{\lambda_i}^{-1}\right) \lambda_i \Bigg) -\frac{s}{2} \ln \sigma_i^2  -\left(1 + \frac{\nu_i + T_i}{2}\right) \ln \sigma^2_i \\
    &\quad - \frac{\nu_i \tau^2_i + \sum_{t=1}^T a_{i,t}\ti{y}_{i,t}^2}{2\sigma^2_i}. 
\end{align*}
Completing square by adding and subtracting the term \begin{align*}  \frac{1}{2 \sigma^2_i} \left(b_i \circ g_i\right)'\left(P_i\circ Q_i + V_{\lambda_i}^{-1}\right)^{-1}\left(b_i \circ g_i\right) \end{align*}
results into 
\begin{align*}
   \ln q\left(\lambda_i, \sigma_{\epsilon_i}^2 \right) &=   C_{\lambda_i,\sigma_{\epsilon_i}^2}  -  \frac{1}{2\sigma_{\epsilon_i}^2}\left(\lambda_i -  \mu_{\lambda_i}\right)'\Sigma_{\lambda_i}^{-1}\left(\lambda_i -  \mu_{\lambda_i}\right) - \frac{s}{2}\ln \sigma_{\epsilon_i}^2\\
    &\quad - \left(1 + \frac{\nu_{\epsilon_i} + T_i}{2}\right) \ln \sigma^2_{\epsilon_i} - \frac{\nu_{\epsilon_i} \tau^2_{\epsilon_i} + \sum_{t=1}^T a_{i,t}\ti{y}_{i,t}^2 - \mu_{\lambda_i}'\Sigma_{\lambda_i}^{-1}\mu_{\lambda_i}}{2\sigma^2_{\epsilon_i}} \\
    &= \ln \mathcal{N}\left(\lambda_i\Big|\mu_{\lambda_i}, \sigma^2_{\epsilon_i} \Sigma_{\lambda_i} \right) +  \ln \text{Scaled Inv-}\chi^2\left(\sigma^2_{\epsilon_i}\Big|\nu_{\epsilon_i} + T_i, \psi_{\epsilon_i}^2 \right), 
\end{align*}
where 
\begin{align*}
       \mu_{\lambda_i} &= \left(P_i \circ Q_i + V_{\lambda_i}^{-1}\right)^{-1}\left(b_i \circ g_i \right), \\
       \Sigma_{\lambda_i} &= \left(P_i \circ Q_i + V_{\lambda_i}^{-1}\right)^{-1},\\
       \psi_{\epsilon_i}^2 &= \frac{1}{\nu_{\epsilon_i} + T_i}\left(\nu_{\epsilon_i}\tau^2_{\epsilon_i} + \sum_{t=1}^T a_{i,t}\ti{y}_{i,t}^2 - \mu_{\lambda_i}'\Sigma_{\lambda_i}^{-1}\mu_{\lambda_i}\right).
\end{align*}

\subsection{Density of $\{\Phi, \Sigma_\ti{u}\}$} \label{sec:A2}
This section is similar to the derivations of S21.\footnote{See Appendix A.3 therein.} By diagonal $\Sigma_u$, we can deduce $q(\Phi, \Sigma_u) = \prod_{j=1}^r q(\phi_j, \sigma^2_{\ti{u}_j})$. Log variational density of $\{\phi_j, \sigma^2_{\ti{u}_j}\}$ is given by 
\begin{align*}
    \ln q\left(\phi_j, \sigma^2_{\ti{u}_j} \right) &= \E{q(F,Z)}{\ln p(Y_A, F, \theta, Z)} =  C_{\phi_j,\sigma^2_{\ti{u}_j}} - \frac{1}{2} T \ln \sigma_{\ti{u}_j}^2 - \sum_{t=1}^T  \frac{1}{2\sigma_{\ti{u}_j}^2}\E{q\left(F,Z\right)}{\left(f_{j,t} - \phi_j'F_{t-1}\right)^2} \\
   &\quad \quad \quad \quad \quad \quad \quad -\frac{s}{2}\ln \sigma_{\ti{u}_j}^2  - \frac{1}{2\sigma^2_{\ti{u}_j}} \phi_j'V_{\phi_j}^{-1}\phi_j - \left(1 + \nu_{\ti{u}_j}/2\right) \ln \sigma^2_{\ti{u}_j} -  \frac{\nu_{\ti{u}_j} \tau^2_{\ti{u}_j}}{2\sigma^2_{\ti{u}_j}}, 
\end{align*} 
where $C_{\phi_j,\sigma^2_{\ti{u}_j}}$ is constant in terms of $\phi_i$ and $\sigma^2_{\ti{u}_j}$.  The expectation term is given by
\begin{align*}
    &\sum_{t=1}^T  \frac{1}{2\sigma_{\ti{u}_j}^2}\E{q\left(F\right)}{\left(f_{j,t} - \phi_j'F_{t-1})\right)^2} \\ 
    &\qquad = \frac{1}{2\sigma^2_{\ti{u}_j}} \left(\sum_{t=1}^T \E{q(F)}{f_{j,t}^2} - 2\phi_j' \sum_{t=1}^T\E{q(F)}{f_{j,t} F_{t-1}'} + \phi_j'\sum_{t=1}^T\E{q(F)}{F_{t-1} F_{t-1}'} \phi_j \right), \numberthis \label{stepA2}
\end{align*}

Inserting \eqref{stepA2} and rearranging terms (e.g. gather all $\lambda_i$-terms):
\begin{align*}
     \ln q\left(\phi_j, \sigma^2_{\ti{u}_j} \right) &=   C_{\phi_j,\sigma^2_{\ti{u}_j}} -  \frac{1}{2\sigma_{\ti{u}_j}^2}\Bigg(- 2\phi_j' \sum_{t=1}^T\E{q(F)}{f_{j,t} F_{t-1}'} + \phi_j'\left(\sum_{t=1}^T\E{q(F)}{F_{t-1} F_{t-1}'} + V_{\phi_j}^{-1}\right)\phi_j \Bigg) \\
    &\quad -\frac{s}{2} \ln \sigma_{\ti{u}_j}^2  -\left(1 + \frac{\nu_{\ti{u}_j} + T}{2}\right) \ln \sigma^2_{\ti{u}_j}
   - \frac{\nu_{\ti{u}_j} \tau^2_{\ti{u}_j} + \sum_{t=1}^T \E{q(F)}{f_{j,t}^2}}{2\sigma^2_{\ti{u}_j}}, 
\end{align*}
and completing square by adding and subtracting the term \begin{align*}  \frac{1}{2 \sigma^2_{\ti{u}_j}} \left(\sum_{t=1}^T\E{q(F)}{f_{j,t} F_{t-1}' }\right)\left(\sum_{t=1}^T\E{q(F)}{F_{t-1} F_{t-1}'} + V_{\phi_j}^{-1}\right)^{-1}\left(\sum_{t=1}^T\E{q(F)}{f_{j,t} F_{t-1}}\right) \end{align*}
yields
\begin{align*}
   \ln q\left(\phi_j, \sigma_{\ti{u}_j}^2 \right) &=   C_{\phi_j,\sigma_{\ti{u}_j}^2}  -  \frac{1}{2\sigma_{\ti{u}_j}^2}\left(\phi_j -  \mu_{\phi_j}\right)'\Sigma_{\phi_j}^{-1}\left(\phi_j -  \mu_{\phi_j}\right) - \frac{s}{2}\ln \sigma_{\ti{u}_j}^2\\
    &\quad - \left(1 + \frac{\nu_{\ti{u}_j} + T}{2}\right) \ln \sigma^2_{\ti{u}_j} - \frac{\nu_{\ti{u}_j} \tau^2_{\ti{u}_j} + \sum_{t=1}^T \E{q(F)}{f_{j,t}^2}- \mu_{\phi_j}'\Sigma_{\phi_j}^{-1}\mu_{\phi_j}}{2\sigma^2_{\ti{u}_j}} \\
    &= \ln \mathcal{N}\left(\phi_j\Big|\mu_{\phi_j}, \sigma^2_{\ti{u}_j} \Sigma_{\phi_j} \right) +  \ln \text{Scaled Inv-}\chi^2\left(\sigma^2_{\ti{u}_j}\Big|\nu_{\ti{u}_j} + T, \psi_{\ti{u}_j}^2 \right) ,
\end{align*}
where 
\begin{align*}
       \mu_{\phi_j} &= \left(\sum_{t=1}^T\E{q(F)}{F_{t-1} F_{t-1}'} + V_{\phi_j}^{-1}\right)^{-1}\left(\sum_{t=1}^T\E{q(F)}{F_{t-1}f_{j,t}}\right), \\
       \Sigma_{\phi_j} &= \left(\sum_{t=1}^T\E{q(F)}{F_{t-1} F_{t-1}'} + V_{\phi_j}^{-1}\right)^{-1},\\
       \psi_{\ti{u}_j}^2 &= \frac{1}{\nu_{\ti{u}_j} + T}\left(\nu_{\ti{u}_j} \tau^2_{\ti{u}_j} + \sum_{t=1}^T \E{q(F)}{f_{j,t}^2}- \mu_{\phi_j}'\Sigma_{\phi_j}^{-1}\mu_{\phi_j}\right).
\end{align*}

\subsection{Density of $F$} \label{sec:A3}
This section shares similarities with derivations of S21.\footnote{See Appendix A.4 therein.} Log variational density of $F$ is given by
\begin{align*}
    \ln q\left(F\right) = \E{q(\theta, Z)}{\ln p(Y_A, F, \theta, Z)}
    &= C_F -\frac{1}{2}\sum_{t=1}^T\E{q\left(\theta, Z\right)}{\left(\ti{y}_t - (Z \circ \Lambda) F_t\right)'A_t \Sigma_\epsilon^{-1} \left(\ti{y}_t - (Z \circ \Lambda) F_t\right)} \\
    &\quad - \frac{1}{2}F_0'V_{F_0}^{-1}F_0 -\frac{1}{2}\sum_{t=1}^T\E{q\left(\theta\right)}{\left(F_t - \widetilde{\Phi} F_{t-1}\right)'\widetilde{\Sigma}_{\ti{u}}^{-1}\left(F_t - \widetilde{\Phi} F_{t-1}\right)}. \numberthis \label{q_F1}
\end{align*}
We need to solve two expectation terms. The first expectation term is given by:
\begin{align*}
    &\E{q\left(\theta, Z\right)}{\left(\ti{y}_t - (Z \circ \Lambda) F_t\right)'A_t \Sigma_\epsilon^{-1} \left(\ti{y}_t - (Z \circ \Lambda) F_t\right)} \\
    &\quad = \te{y}_t'A_t\E{q\left(\theta\right)}{\Sigma_\epsilon^{-1}}\ti{y}_t -  2F_t'\E{q\left(\theta,Z\right)}{(Z \circ \Lambda)'A_t\Sigma_\epsilon^{-1}}\ti{y}_t + F_t'\E{q\left(\theta, Z\right)}{(Z \circ \Lambda)'A_t\Sigma_\epsilon^{-1}(Z \circ \Lambda)} F_t,
\end{align*}
where $\E{q\left(\theta\right)}{\Sigma_\epsilon^{-1}} = \Psi^{-1}_\epsilon$ and by law of iterated expectations
\begin{align*}
    \E{q\left(\theta,Z\right)}{(Z \circ \Lambda)'A_t\Sigma_\epsilon^{-1}} = \E{q(\Sigma_\epsilon)}{(\E{q\left(Z\right)}{Z} \circ \E{q\left(\Lambda|\Sigma_\epsilon\right))}{\Lambda})'A_t\Sigma_\epsilon^{-1}} = \underset{Z \circ \Lambda}{M} 'A_t\Psi_\epsilon^{-1}.
\end{align*}
To solve $\E{q\left(\theta,Z\right)}{(Z \circ \Lambda)'A_t\Sigma_\epsilon^{-1}}\ti{y}_t + F_t'\E{q\left(\theta, Z\right)}{(Z \circ \Lambda)'A_t\Sigma_\epsilon^{-1}(Z \circ \Lambda)}$ we note that 

\begin{align*}
&(Z \circ \Lambda)'A_t\Sigma_\epsilon^{-1}(Z \circ \Lambda) = \\
&\qquad \begin{bmatrix} \sum_{i=1}^n \frac{a_{i,t}}{\sigma^2_{\epsilon_i}} z^2_{i,1}\lambda^2_{i,1} & \sum_{i=1}^n \frac{a_{i,t}}{\sigma^2_{\epsilon_i}} z_{i,1}z_{i,2}\lambda_{i,1} \lambda_{i,2} & \dots & \sum_{i=1}^n \frac{a_{i,t}}{\sigma^2_{\epsilon_i}} z_{i,1}z_{i,n}\lambda_{i,1} \lambda_{i,n} \\
\sum_{i=1}^n \frac{a_{i,t}}{\sigma^2_{\epsilon_i}} z_{i,2}z_{i,1}\lambda_{i,2} \lambda_{i,1} & \sum_{i=1}^n \frac{a_{i,t}}{\sigma^2_{\epsilon_i}} z^2_{i,2}\lambda^2_{i,2} & \dots & \sum_{i=1}^n \frac{a_{i,t}}{\sigma^2_{\epsilon_i}} z_{i,2}z_{i,n} \lambda_{i,2} \lambda_{i,n} \\
\vdots & \vdots & \ddots & \vdots \\ 
\sum_{i=1}^n \frac{a_{i,t}}{\sigma^2_{\epsilon_i}} z_{i,n}z_{i,1} \lambda_{i,n} \lambda_{i,1} & \sum_{i=1}^n \frac{a_{i,t}}{\sigma^2_{\epsilon_i}} z_{i,n}z_{i,2}\lambda_{i,n} \lambda_{i,2} & \dots & \sum_{i=1}^n \frac{a_{i,t}}{\sigma^2_{\epsilon_i}} z^2_{i,n}\lambda^2_{i,n}\end{bmatrix}. \numberthis \label{LambdaSquare}
\end{align*}

Taking the expecation of an element in \eqref{LambdaSquare} gives
\begin{align*}
    &\E{q(\theta,Z)}{\sum_{i=1}^n \frac{a_{i,t}}{\sigma^2_{\epsilon_i}} z_{i,k}z_{i,m}\lambda_{i,k}\lambda_{i,m}} = \E{q(\Sigma_\epsilon)}{\sum_{i=1}^n\frac{a_{i,t}}{\sigma^2_i} \E{q(z_i)}{z_{i,k}z_{i,m}}\E{q(\lambda_i|\sigma^2_{\epsilon_i})}{\lambda_{i,k}\lambda_{i,m}}} \\
    &\qquad = \E{q(\Sigma_\epsilon)}{\sum_{i=1}^n\frac{a_{i,t}}{\sigma^2_{\epsilon_i}}\left(b_{i,k}b_{i,m} + \text{Cov}_{q(z_i)}(z_{i,k},z_{i,m})\right)\left(\mu_{\lambda_{i,k}} \mu_{\lambda_{i,m}} + \sigma^2_{\epsilon_i} [\Sigma_{\lambda_i}]_{k,m}\right)} \\
    &\qquad = \underbrace{\sum_{i=1}^n\frac{a_{i,t}}{\psi^2_{\epsilon_i}} b_{i,k}b_{i,m}\mu_{\lambda_{i,k}} \mu_{\lambda_{i,m}}}_{\text{first term}} + \underbrace{\sum_{i=1}^n a_{i,t} [P_i]_{k,m} [\Sigma_{\lambda_i}]_{k,m}}_{\text{second term}} +  \underbrace{\sum_{i=1}^n\frac{a_{i,t}}{\psi^2_{\epsilon_i}} \text{Cov}_{q(z_i)}(z_{i,k},z_{i,m}) \mu_{\lambda_{i,k}} \mu_{\lambda_{i,m}}}_{\text{third term}} \numberthis \label{lambdaelement}
\end{align*}

where we have used $[P_i]_{k,m} = b_{i,k}b_{i,m} + \text{Cov}_{q(z_i)}(z_{i,k},z_{i,m})$. We note that \begin{itemize} 
    \item the first term is the $(k,m)$-element of $(B \circ M_\Lambda)'A_t\Psi_\epsilon^{-1}(B \circ M_\Lambda) = \underset{Z \circ \Lambda}{M} 'A_t\Psi_\epsilon^{-1}\underset{Z \circ \Lambda}{M}$, 
    \item the second term is the $(k,m)$-element of $\sum_{i=1}^n a_{i,t}\left(P_i \circ \Sigma_{\lambda_i}\right)$.
\end{itemize} As $z_{i,k}$ is independent Bernoulli distributed (see Appendix \ref{sec.A4}), the third term is given by
\begin{align*}
    \sum_{i=1}^n\frac{a_{i,t}}{\psi^2_{\epsilon_i}} \text{Cov}_{q(z_i)}(z_{i,k},z_{i,m}) \mu_{\lambda_{i,k}} \mu_{\lambda_{i,m}} = \begin{cases} \sum_{i=1}^n\frac{a_{i,t}}{\psi^2_{\epsilon_i}} b_{i,k}\left(1-b_{i,k}\right)\mu_{\lambda_{i,k}}^2 \quad &\text{ if } k=m \\
    0 \quad &\text{ if } k \neq m\end{cases}.
\end{align*}
We thus know that it is a the third term is an element from a diagonal matrix. Note that $\frac{1}{\psi^2_{\epsilon_i}}b_{i,k}\left(1-b_{i,k}\right)\mu_{\lambda_{i,k}}^2$ is the $(i,k)$-element of matrix 
\begin{align*}
    W =  \Psi_\epsilon^{-1}(B \circ (\mathbf{1}_{n \times s} - B) \circ M_\Lambda \circ M_\Lambda).
\end{align*}
Then $\sum_{i=1}^n\frac{a_{i,t}}{\psi^2_{\epsilon_i}} b_{i,k}\left(1-b_{i,k}\right)\mu_{\lambda_{i,k}}^2$ is the sum of column $k$ in $W$, over rows corresponding to available elements. From this we can deduce that the third term is the $(k,m)$-element of 
\begin{align*}
    \sum_{i=1}^n a_{i,t} \text{diag}\left(w_i\right),
\end{align*} 
where $w_i$ is the $i$th row of $W$. Consequently, by \eqref{LambdaSquare}-\eqref{lambdaelement}, we have the expression
\begin{align*}
    \E{q\left(\theta, Z\right)}{(Z \circ \Lambda)'A_t\Sigma_\epsilon^{-1}(Z \circ \Lambda)} = \underset{Z \circ \Lambda}{M} 'A_t\Psi_\epsilon^{-1}\underset{Z \circ \Lambda}{M} + \sum_{i=1}^n a_{i,t}\Big(\left(P_i \circ \Sigma_{\lambda_i}\right) +  \text{diag}\left(w_i\right)\Big). \numberthis \label{E1qF}
\end{align*}
Now, we turn back to \eqref{q_F1}. The second expectation term is given by
\begin{align*} 
   &\E{q\left(\theta\right)}{\left(F_t - \widetilde{\Phi} F_{t-1}\right)'\widetilde{\Sigma}_{\ti{u}}^{-1}\left(F_t - \widetilde{\Phi} F_{t-1}\right)} \\
    \quad &= F_t'\E{q\left(\Sigma_{\ti{u}}\right)}{\widetilde{\Sigma}_{\ti{u}}^{-1}}F_t - F_t'\E{q\left(\theta\right)}{\widetilde{\Sigma}_{\ti{u}}^{-1}\widetilde{\Phi}} F_{t-1} - F_{t-1}'\E{q\left(\theta\right)}{\widetilde{\Phi}'\widetilde{\Sigma}_{\ti{u}}^{-1}} F_t + F_{t-1}'\E{q\left(\theta\right)}{\widetilde{\Phi}'\widetilde{\Sigma}_{\ti{u}}^{-1}\widetilde{\Phi}}F_{t-1}
\end{align*}

where 
\begin{align} 
     \E{q\left(\Phi|\Sigma_\ti{u}\right)}{\widetilde{\Phi}} \equiv \widetilde{M}_{\Phi} =  \left[
    \begin{array}{cc}
      \multicolumn{2}{c}{M_\Phi}  \\ \hdashline[2pt/2pt]
       \multicolumn{1}{c}{\underset{rp \times rp}{I}} & \multicolumn{1}{;{2pt/2pt}c}{\underset{rp \times r}{0}}
    \end{array}\right], \quad      \E{q\left(\Sigma_\ti{u}\right)}{\widetilde{\Sigma}_\ti{u}^{-1}} \equiv \widetilde{\Psi}^{-1}_\ti{u} = S \Psi^{-1}_\ti{u}S' =  \begin{bmatrix} \Psi^{-1}_\ti{u} & \underset{r \times rp}{0} \\
    \underset{rp \times r}{0} & \underset{rp \times rp}{0}
    \end{bmatrix}. \label{linearEPhi}\end{align}

To solve $\E{q\left(\theta\right)}{\widetilde{\Phi}'\widetilde{\Sigma}_{\ti{u}}^{-1}\widetilde{\Phi}}$ we note that 
\begin{align*}
    \widetilde{\Phi}'\widetilde{\Sigma}_{\ti{u}}^{-1}\widetilde{\Phi} = \Phi' \Sigma_\ti{u}^{-1} \Phi = \begin{bmatrix} \sum_{j=1}^r \frac{\phi^2_{j,1}}{\sigma^2_{\ti{u}_j}} & \sum_{j=1}^r \frac{\phi_{j,1} \phi_{j,2}}{\sigma^2_{\ti{u}_j}} & \dots & \sum_{j=1}^r \frac{\phi_{j,1} \phi_{j,r}}{\sigma^2_{\ti{u}_j}}  \\
\sum_{j=1}^r \frac{\phi_{j,2} \phi_{j,1}}{\sigma^2_{\ti{u}_j}}  & \sum_{j=1}^r \frac{\phi^2_{j,2}}{\sigma^2_{\ti{u}_j}} & \dots & \sum_{j=1}^r \frac{\phi_{j,2} \phi_{j,r}}{\sigma^2_{\ti{u}_j}}  \\
\vdots & \vdots & \ddots & \vdots \\ 
\sum_{j=1}^r \frac{\phi_{j,r} \phi_{j,1}}{\sigma^2_{\ti{u}_j}}  & \sum_{j=1}^r \frac{\phi_{j,r} \phi_{j,2}}{\sigma^2_{\ti{u}_j}} & \dots & \sum_{j=1}^r \frac{\phi^2_{j,r}}{\sigma^2_{\ti{u}_j}} \end{bmatrix},
\end{align*}
where the expectation of a single element is given by
\begin{align*}
    \E{\theta}{\sum_{j=1}^r \frac{\phi_{j,k} \phi_{j,m}}{\sigma^2_{\ti{u}_j}}} &= \sum_{j=1}^r \E{q(\Sigma_\ti{u})}{\frac{\E{q(\Phi|\Sigma_\ti{u})}{\phi_{j,k} \phi_{j,m}}}{\sigma^2_{\ti{u}_j}}} = \sum_{j=1}^r \E{q(\Sigma_\ti{u})}{\frac{\mu_{\phi_{j,k}} \mu_{\phi_{j,m}} + \sigma^2_{\ti{u}_j}[\Sigma_{\phi_j}]_{k,m} }{\sigma^2_{\ti{u}_j}}} \\ 
    &= \sum_{j=1}^r \frac{\mu_{\phi_{j,k}} \mu_{\phi_{j,m}}}{\psi^2_{\ti{u}_j}} +  + \sum_{j=1}^r [\Sigma_{\phi_j}]_{k,m},
\end{align*}
which is the $(k,m)$-element of
\begin{align*}
    \E{q\left(\theta\right)}{\widetilde{\Phi}'\widetilde{\Sigma}_{\ti{u}}^{-1}\widetilde{\Phi}} = \widetilde{M}'_\Phi \widetilde{\Sigma}_\ti{u}^{-1}  \widetilde{M}_\Phi + \sum_{j=1}^r \Sigma_{\phi_j}. 
\end{align*}
Inserting all solved expectations in \eqref{q_F1} gives
\begin{align*}
    \ln q(F) &= C_F -\frac{1}{2}\sum_{t=1}^T\Bigg(\ti{y}_t'A_t\Psi^{-1}\ti{y}_t -  2F_t'\underset{Z \circ \Lambda}{M} 'A_t\Psi_\epsilon^{-1}\ti{y}_t + F_t'\underset{Z \circ \Lambda}{M}'A_t\Psi_\epsilon^{-1}\underset{Z \circ \Lambda}{M}  \\
    &\qquad \qquad \qquad \qquad \qquad \qquad +  F_t'\left(\sum_{i=1}^n a_{i,t}\Big(\left(P_i \circ \Sigma_{\lambda_i}\right) +  \text{diag}\left(w_i\right)\Big)\right)F_t\Bigg) \\
    &\quad - \frac{1}{2}F_0'V_{F_0}^{-1}F_0 -\frac{1}{2}\sum_{t=1}^T\Bigg(F_t'\widetilde{\Psi}^{-1}_\ti{u}F_t - F_t'\widetilde{\Psi}^{-1}_\ti{u}\widetilde{M}_\Phi F_{t-1} - F_{t-1}'\widetilde{M}_\Phi'\widetilde{\Psi}^{-1}_\ti{u} F_t \\
    &\qquad \qquad \qquad \qquad \qquad \qquad + F_{t-1}'\left(\widetilde{M}_\Phi'\widetilde{\Psi}^{-1}_\ti{u}\widetilde{M}_\Phi + \sum_{j=1}^r\Sigma_{\phi_j}\right)F_{t-1}\Bigg). 
\end{align*}
Rearranging terms
\begin{align*}
    \ln q(F) &= C_F -\frac{1}{2}\sum_{t=1}^T\left(\ti{y}_t'A_t\Psi_\epsilon^{-1}\ti{y}_t -  2F_t'\underset{Z \circ \Lambda}{M} 'A_t\Psi_\epsilon^{-1}\ti{y}_t + F_t'\underset{Z \circ \Lambda}{M}'A_t\Psi_\epsilon^{-1}\underset{Z \circ \Lambda}{M}  \right) \\
         &- \frac{1}{2}F_0'V_{F_0}^{-1}F_0 -\frac{1}{2}\sum_{t=1}^T\left(F_t'\widetilde{\Psi}^{-1}_\ti{u}F_t - F_t'\widetilde{\Psi}^{-1}_\ti{u}\widetilde{M}_\Phi F_{t-1} - F_{t-1}'\widetilde{M}_\Phi'\widetilde{\Psi}^{-1}_\ti{u} F_t + F_{t-1}'\widetilde{M}_\Phi'\widetilde{\Psi}^{-1}_\ti{u}\widetilde{M}_\Phi F_{t-1}\right) \\
    &+ \sum_{t=1}^T F_t'\left(\sum_{i=1}^n a_{i,t}\Big(\left(P_i \circ \Sigma_{\lambda_i}\right) +  \text{diag}\left(k_i\right)\Big)\right)F_t + \sum_{t=0}^{T-1}F_t'\left(\sum_{j=1}^r \Sigma_{\phi_j}\right) F_t.
\end{align*}
Completing squares and rearranging remaining terms after time index
\begin{align*}
     \ln q(F) &= C_F -\frac{1}{2}\sum_{t=1}^T\left(\ti{y}_t - \underset{Z \circ \Lambda}{M} F_t\right)'A_t \Psi_\epsilon^{-1}\left(\ti{y}_t - \underset{Z \circ \Lambda}{M} F_t\right) -\frac{1}{2}\sum_{t=1}^T \left(F_t - \widetilde{M}_\Phi F_{t-1}\right)'\widetilde{\Psi}^{-1}_\ti{u}\left(F_t - \widetilde{M}_\Phi F_{t-1}\right) \\ 
    &\quad- \frac{1}{2}F_0'\left(V_{F_0}^{-1} + \sum_{j=1}^r\Sigma_{\phi_j}\right)F_0  - \frac{1}{2}\sum_{t=1}^{T-1}F_t'\left(\sum_{i=1}^n a_{i,t}\Big(\left(P_i \circ \Sigma_{\lambda_i}\right) +  \text{diag}\left(w_i\right)\Big) + \sum_{j=1}^r \Sigma_{\phi_j} \right)F_t \\ &\quad - \frac{1}{2}F_T'\left(\sum_{i=1}^n a_{i,t}\Big(\left(P_i \circ \Sigma_{\lambda_i}\right) +  \text{diag}\left(w_i\right)\Big)\right)F_T,
\end{align*}
and inserting the last two terms in the first square by augmenting vectors and matrices yield
\begin{align*}
    \ln q(F) &= C_F -\frac{1}{2}\sum_{t=1}^T\left(\begin{bmatrix} \ti{y}_t \\ 0_{s \times 1} \end{bmatrix} - \begin{bmatrix} \underset{Z \circ \Lambda}{M} \\[7pt] I_{s} \end{bmatrix}  F_t\right)'\begin{bmatrix}A_t\Psi_\epsilon^{-1} & 0_{n \times s} \\ 0_{s \times n} & \Sigma^{(\theta,Z)}_t \end{bmatrix}\left(\begin{bmatrix} \ti{y}_t \\ 0_{s \times 1} \end{bmatrix} - \begin{bmatrix} \underset{Z \circ \Lambda}{M} \\[7pt] I_{s} \end{bmatrix}  F_t\right)\\
    &\quad - \frac{1}{2}\sum_{t=1}^T \left(F_t - \widetilde{M}_\Phi F_{t-1}\right)'\widetilde{\Psi}_\ti{u}^{-1}\left(F_t - \widetilde{M}_\Phi F_{t-1}\right) -  \frac{1}{2}F_0'\left(V_{F_0}^{-1} + \sum_{j=1}^r \Sigma_{\phi_j}\right)F_0, \numberthis \label{q_F3}
\end{align*}
where 
\begin{align*}
           \Sigma^{\theta Z}_t = \begin{cases}  \sum_{i=1}^n a_{i,t}\Big(\left(P_i \circ \Sigma_{\lambda_i}\right) +  \text{diag}\left(w_i\right)\Big) + \sum_{j=1}^r \Sigma_{\phi_j}\quad &\text{ if } t=1, ..., T-1 \\ \\
    \sum_{i=1}^n a_{i,t}\Big(\left(P_i \circ \Sigma_{\lambda_i}\right) +  \text{diag}\left(w_i\right)\Big) \quad &\text{ if } t=T \end{cases}.
\end{align*}

\subsection{Density of $Z$} \label{sec.A4}
By definition $q(Z) = \prod_{i=1}^n \prod_{k=1}^s q(z_{i,k})$. The log variational density of $z_{i,k}$ is given by
\begin{align*}
    \ln q(z_{i,k}) = C_{z_{i,k}} - \E{q(F,\theta, Z\backslash z_{i,k})}{\sum_{t=1}^T \frac{a_{i,t}}{2\sigma_{\epsilon_i}^2}\left(\ti{y}_{i,t} - (z_i \circ \lambda_i)'F_t\right)^2} + z_{i.k}\text{logit}(\beta_{i,k}), \numberthis \label{qzik}
\end{align*}
where $C_{z_{i,k}}$ is constant in terms of $z_{i,k}$. Let us define
\begin{align*}
\mathcal{E}_{z_{i,k}}  \equiv  \E{q(F,\theta, Z\backslash z_{i,k})}{\sum_{t=1}^T \frac{a_{i,t}}{2\sigma_{\epsilon_i}^2}\left(\ti{y}_{i,t} - (z_i \circ \lambda_i)'F_t\right)^2},
\end{align*} 
which is given by
 \begin{align*}
     &\mathcal{E}_{z_{i,k}}  = \E{q(F,\theta, Z\backslash z_{i,k})}{\sum_{t=1}^T \frac{a_{i,t}}{2\sigma_{\epsilon_i}^2}\left(\ti{y}_{i,t} - \sum_{m=1}^s z_{i,m}\lambda_{i,m} F_{m,t}\right)^2}  \\
     =& \sum_{t=1}^T \frac{a_{i,t}}{2\sigma_{\epsilon_i}^2} \ti{y}^2_{i,t} - 2\E{q(F,\theta, Z\backslash z_{i,k})}{\sum_{t=1}^T \frac{a_{i,t}}{2\sigma_{\epsilon_i}^2}\ti{y}_{i,t}\left(\sum_{m=1}^s z_{i,m}\lambda_{i,m} F_{m,t}\right)} \\
     &\quad + \E{q(F,\theta, Z\backslash z_{i,k})}{\sum_{t=1}^T \frac{a_{i,t}}{2\sigma_{\epsilon_i}^2}\left(\sum_{m=1}^s z_{i,m}\lambda_{i,m} F_{m,t}\right)^2},
\end{align*}
  where we have the square term
 \begin{align*}
     \left(\sum_{m=1}^s z_{i,m}\lambda_{i,m} F_{m,t}\right)^2 = \left(\sum_{m=1}^s z_{i,m}^2\lambda_{i,m}^2 F_{m,t}^2 + 2\sum_{\ell=2}^s\sum_{m=1}^{\ell-1} z_{i,m}z_{i,\ell}\lambda_{i,m}\lambda_{i,\ell}F_{m,t}F_{\ell,t}\right).
 \end{align*}
We know that $z_{i,m}^2 = z_{i,m}$. Additionally, every term which does not include $z_{i,k}$ can be put into a constant $C^{(1)}_{z_{i,k}}$. Consequently, 
 \begin{align*}
     \mathcal{E}_{z_{i,k}} &= C^{(1)}_{z_{i,k}} - \E{q(F,\theta)}{\sum_{t=1}^T \frac{a_{i,t}}{\sigma_{\epsilon_i}^2}\ti{y}_{i,t} z_{i,k}\lambda_{i,k} F_{k,t}} \\
     &\quad + \E{q(F,\theta, Z\backslash z_{i,k})}{\sum_{t=1}^T \frac{a_{i,t}}{2\sigma_i^2}\left(z_{i,j}\lambda_{i,k}^2 F_{k,t}^2 + 2 z_{k,j} \sum_{m \neq k} z_{i,m} \lambda_{i,k} \lambda_{i,m} F_{k,t} F_{m,t}\right)}.
\end{align*}
Factorizing $z_{i,k}$ from each term and rearranging expectations
\begin{align*}
      \mathcal{E}_{z_{i,k}} &= C^{(1)}_{z_{i,k}} + z_{i,k}\left(-\E{q(\theta)}{\frac{\lambda_{i,k}}{\sigma_{\epsilon_i}^2}}\left(\sum_{t=1}^T a_{i,t}\ti{y}_{i,t} \E{q(F)}{F_{k,t}}\right) + \frac{1}{2}\E{q(\theta)}{\frac{\lambda_{i,k}^2}{\sigma_{\epsilon_i}^2}} \left(\sum_{t=1}^T a_{i,t}\E{q(F)}{F_{k,t}^2}\right) \right.\\
     &\qquad \qquad \qquad \qquad \qquad \qquad \qquad  + \left.\sum_{m \neq k} \E{q(z_{i,m})}{z_{i,m}} \E{q(\theta)}{\frac{\lambda_{i,k} \lambda_{i,m}}{\sigma_{\epsilon_i}^2}}\left(\sum_{t=1}^T a_{i,t}\E{q(F)}{F_{k,t} F_{m,t}}\right)\right).
\end{align*}
 
We observe that $g_{i,k} = \sum_{t=1}^T a_{i,t}\ti{y}_{i,t} \E{q(F)}{F_{k,t}}$ is the $k$th element of $g_i$ and $[Q_i]_{k,m} = \sum_{t=1}^T a_{i,t}\E{q(F)}{F_{k,t} F_{m,t}}$. Also, let's denote
\begin{align*}
    R_i = \E{q(\theta)}{\frac{1}{\sigma^2_i}\lambda_i \lambda_i'}.
\end{align*}
Then we have $[R_i]_{k,m} = \E{q(\theta)}{\frac{\lambda_{i,k} \lambda_{i,m}}{\sigma^2_{\epsilon_i}}}$. We can rewrite
 \begin{align*}
     \mathcal{E}_{z_{i,k}} = C^{(1)}_{z_{i,k}} + z_{i,k}\left(-\frac{\mu_{\lambda_{i,k}}g_{i,k}}{\psi_{\epsilon_i}^2} + \frac{1}{2}[R_i]_{k,k} [Q_i]_{k,k} + \sum_{m \neq k} b_{i,m} [R_i]_{k,m}[Q_i]_{k,m}\right). \numberthis \label{Ezik}
 \end{align*}

Denote
\begin{align*}
    \gamma_{i,k} = \frac{\mu_{\lambda_{i,k}}g_{i,k}}{\psi_{\epsilon_i}^2} - \frac{1}{2}[R_i]_{k,k} [Q_i]_{k,k} - \sum_{m \neq k} b_{i,m} [R_i]_{k,m}[Q_i]_{k,m}
\end{align*}
and inserting \eqref{Ezik} in \eqref{qzik} yields
\begin{align*}
     \ln q(z_{i,k}) = C_{z_{i,k}}^{(2)} + z_{i,k}\left(\gamma_{i,k} + \text{logit}(\beta_{i,k})\right),
\end{align*}
where $C_{z_{i,k}}^{(2)}$ is constant in terms of $z_{i,k}$. We can see that
\begin{align*}
    \gamma_{i,k} + \text{logit}(\beta_{i,k}) &= \ln\left(\exp\left\{  \gamma_{i,k} + \text{logit}(\beta_{i,k})\right\}\right) = \text{logit}\left(\frac{1}{1 + \exp\{-\gamma_{i,k}- \text{logit}(\beta_{i,k})\}}\right) 
\end{align*} 
using arithmetic fact $c = d/(1 - d) \Rightarrow d = 1/(1+c^{-1})$. Conclusively

 \begin{align*}
    \ln q(z_{i,k}) = C_{z_{i,k}}^{(2)} + z_{i.k}\text{logit}(b_{i,k}) = \ln\text{Be}\left(z_{i,k}|b_{i,k}\right),
\end{align*}
where 
\begin{align*}
    b_{i,k} = \text{expit}(\gamma_{i,k} + \text{logit}(\beta_{i,k})).
\end{align*}

\subsection{Evidence lower bound} \label{sec:A5}
The results in this section follows closely derivations of S21. \footnote{See Appendix A.5 therein.} The evidence lower bound is given by 
\begin{align}
    \text{ELBO} = \E{q}{\ln p(\Omega, F, \theta, Z)} - \E{q}{\ln q(F)} - \E{q}{\ln q(\theta)} - \E{q}{\ln q(Z)}. \label{ELBO1}
\end{align} 

Introducing  
\begin{align*}
    \mathcal{E}_F &= -\frac{1}{2}\sum_{t=1}^T\E{q}{\left(\ti{y}_t - (Z \circ \Lambda) F_t\right)'A_t \Sigma_\epsilon^{-1} \left(\ti{y}_t - (Z \circ \Lambda) F_t\right)} \\
    &\quad - \frac{1}{2}\E{q}{F_0'V_{F_0}^{-1}F_0} -\frac{1}{2}\sum_{t=1}^T\E{q}{\left(F_t - \widetilde{\Phi} F_{t-1}\right)'\left(F_t - \widetilde{\Phi} F_{t-1}\right)}
\end{align*}
then 
\begin{align*}
    \E{q}{\ln p(\Omega, F, \theta, Z)} &= \mathcal{E}_F -\frac{1}{2} \sum_{i=1}^n T_i \ln 2\pi -\frac{1}{2} \sum_{i=1}^n T_i \E{q}{\ln \sigma_{\epsilon_i}^2} -\frac{1}{2} \ln \det\left(V_{F_0}\right) -\frac{rT + s}{2} \ln 2\pi  \\
    &\quad- \frac{T}{2} \sum_{j=1}^r \E{q}{\ln \sigma_{\ti{u}_j}^2}  +\E{q}{\ln p\left(\theta\right)} + \E{q}{\ln p(Z)}, \numberthis \label{Ejoint}  \\
   \E{q}{\ln q\left(F\right)} &= C_{F} + \mathcal{E}_F. \numberthis \label{EqF}
\end{align*}

By property of Scaled-inverse-chi-square distribution we have expectation
\begin{align*}
    \E{q}{\ln \sigma_{\epsilon_i}^2} &= \ln \left((\nu_{\epsilon_i} + T_i) \psi^2_{\epsilon_i} \right) - \ln 2 - \varphi\left(\frac{\nu_{\epsilon_i} + T_i}{2}\right), \numberthis \label{Esigma}
\end{align*}
with a corresponding expression for $\E{q}{\ln \sigma_{u_j}^2}$. Inserting \eqref{Ejoint}-\eqref{Esigma} in \eqref{ELBO1} gives

\begin{align*}
    \text{ELBO} &= -C_F -\frac{1}{2} \sum_{i=1}^n T_i \ln \pi  -\frac{1}{2} \sum_{i=1}^n T_i \ln \left((\nu_{\epsilon_i} + T_i)  \psi^2_{\epsilon_i}\right)  -\frac{T}{2} \sum_{j=1}^r \ln \left(\frac{\nu_{\ti{u}_j} + T}{2}\right) - \frac{T}{2} \sum_{j=1}^r \ln \psi^2_{\ti{u}_j}  \\
    &\quad +  \sum_{i=1}^n \frac{T_i}{2} \varphi\left(\frac{\nu_{\epsilon_i} + T_i}{2}\right) + \sum_{j=1}^r \frac{T}{2}\varphi\left(\frac{\nu_{\ti{u}_j} + T}{2}\right)-\frac{1}{2} \ln \left(\frac{\det\left(V_{F_0}\right)}{\det\left(\Sigma_{F_0}\right)}\right) \\ 
    &\quad - \Big(\E{q}{\ln q(\theta)} - \E{q}{\ln p\left(\theta\right)}\Big) - \Big(\E{q}{\ln q(Z)} - \E{q}{\ln p\left(Z\right)}\Big). \numberthis \label{ELBO2}
\end{align*}

From derivation of S21 we get

\begin{align*}
    C_{F} &= -\frac{rT + s}{2} \ln 2\pi - \frac{T}{2} \sum_{j=1}^r \ln \psi^2_{\ti{u}_j} 
    - \frac{1}{2} \ln \det\left(\Sigma_{F_0}\right) + \frac{1}{2} \ln \left(\frac{\det\left(G_t\right)}{\det\left(H^\star_t\right)} \right) \\ 
    &\quad + \frac{1}{2}\sum_{t=1}^T {\widehat{\epsilon^\star}}'_t G_t^{-1} \widehat{\epsilon^\star}_t + \frac{1}{2} \sum_{i=1}^n {e_t^S}' S^{A'}_t \Psi^{-1}_\epsilon S^A_t e_t^S  + \frac{1}{2} \sum_{i=1}^n  {\ti{y}^\star_t}' \Sigma_t^{\theta Z} \ti{y}^\star_t, \numberthis \label{C_F}
\end{align*} 

where 
\begin{align*}
\widehat{\epsilon^\star}_t &= \ti{y}_t^\star - \E{q(F)}{F_t|Y^\star_{t-1}},  \\
G_t &= \text{Cov}_{q(F)}\left[F_t|Y^\star_{t-1}\right] + H^\star_t, \\
    e_t^S &= S^A_t\ti{y}_t - S^A_t\underset{Z \circ \Lambda}{M} \ti{y}_t^\star,  
\end{align*} 

in which $\E{q(F)}{F_t|Y^\star_{t-1}}$ and $\text{Cov}_{q(F)}\left[F_t|Y^\star_{t-1}\right]$ are the predictive state mean and covariance, respectively, from Kalman filter over collapsed model \eqref{Fss2}-\eqref{FssC}. $S^A_t$ is a selection matrix, formed by the rows of a identity matrix corresponding to the available elements in $\ti{y}_t$.

We note that the two last parenthesis in ELBO are KL-divergences between variational densities and prior densities, given by
\begin{align*} 
\E{q}{\ln q(\theta)} - \E{q}{\ln p\left(\theta\right)} &= \sum_{i=1}^n\E{q\left(\sigma_{\epsilon_i}^2\right)}{\text{KL}\left(q\left(\lambda_i|\sigma_{\epsilon_i}^2\right)\Big|\Big|p\left(\lambda_i|\sigma_{\epsilon_i}^2\right)\right)} + \sum_{i=1}^n\text{KL}\left(q\left(\sigma_{\epsilon_i}^2\right)\Big|\Big|p\left(\sigma_{\epsilon_i}^2\right)\right) \\ 
&\quad+ \sum_{j=1}^r\E{q\left(\sigma_{\ti{u}_j}^2\right)}{\text{KL}\left(q\left(\phi_j|\sigma_{\ti{u}_j}^2\right)\Big|\Big|p\left(\phi_j|\sigma_{\ti{u}_j}^2\right)\right)} + \sum_{i=1}^n\text{KL}\left(q\left(\sigma_{\ti{u}_j}^2\right)\Big|\Big|p\left(\sigma_{\ti{u}_j}^2\right)\right), \\ 
\E{q}{\ln q(Z)} - \E{q}{\ln p\left(Z\right)} &=  \sum_{i=1}^n \sum_{k=1}^s \text{KL}\left(q(z_{i,k})\Big|\Big|p(z_{i,k})\right).
\end{align*}
We have Gaussian KL-divergences
\begin{align}
    \E{q\left(\sigma_{\epsilon_i}^2\right)}{\text{KL}\left(q\left(\lambda_i|\sigma_{\epsilon_i}^2\right)\Big|\Big|p\left(\lambda_i|\sigma_{\epsilon_i}^2\right)\right)} &= -\frac{s}{2} + \frac{1}{2}\Tr\left(V_{\lambda_i}^{-1}\Sigma_{\lambda_i}\right) + \frac{1}{2}{\mu_{\lambda_i}}'V_{\lambda_i}^{-1}\mu_{\lambda_i} +  \frac{1}{2}\ln \left(\frac{\det\left(V_{\lambda_i}\right)}{\det\left(\Sigma_{\lambda_i}\right)}\right), \label{KLlambda}
\end{align} 
with a corresponding expression for $q(\phi_j|\sigma^2_u)$, and 
scaled-Inverse-chi-square KL-divergences
\begin{align*}
    \text{KL}\left(q\left(\sigma_{\epsilon_i}^2\right)\Big|\Big|p\left(\sigma_{\epsilon_i}^2\right)\right) &= \frac{T_i}{2} \varphi\left(\frac{\nu_{\epsilon_i} + T_i}{2}\right) - \ln \Gamma\left(\frac{\nu_{\epsilon_i} + T_i}{2}\right) + \ln \Gamma\left(\frac{\nu_{\epsilon_i}}{2}\right) + \frac{\nu_{\epsilon_i}}{2}\ln\left( (\nu_{\epsilon_i} + T_i) \psi^2_{\epsilon_i}\right) \\ &\quad- \frac{\nu_{\epsilon_i}}{2}\ln \nu_{\epsilon_i} \tau^2_{\epsilon_i}
     + \frac{\nu_{\epsilon_i}\tau^2_{\epsilon_i}}{2\psi^2_{\epsilon_i}} - \frac{\nu_{\epsilon_i} + T_i}{2}, \numberthis \label{KLsigma} 
\end{align*}
with a corresponding expression for $q(\sigma^2_u)$. Lastly, we have Bernoulli KL-divergences
\begin{align*}
    \text{KL}\left(q(z_{i,k})\Big|\Big|p(z_{i,k})\right) = b_{i,k} \ln \left(\frac{b_{i,k}}{\beta_{i,k}}\right) + (1-b_{i,k})\ln\left( \frac{1-b_{i,k}}{1-\beta_{i,k}}\right). \numberthis \label{KLz}
\end{align*}
We get the full expression of ELBO by insert the constant \eqref{C_F} and KL-divergences  \eqref{KLlambda}-\eqref{KLz} in \eqref{ELBO2}:
\begin{align}
    \text{ELBO} = \{F\text{-terms}\} + \{\Lambda\text{-terms}\} + \{\Phi\text{-terms}\} + \{\Sigma_\epsilon\text{-terms}\} +  \{\Sigma_\ti{u}\text{-terms}\} + \{Z\text{-terms}\}, 
\end{align}
where 
\begin{align*}
    \{F\text{-terms}\} &= -\frac{1}{2}\sum_{i=1}^n T_i \ln \pi - \frac{1}{2}\ln \left(\frac{\det\left(V_{F_0}\right)}{\det\left(\Sigma_{F_0}\right)}\right) - \frac{1}{2} \ln \left(\frac{\det\left(G_t\right)}{\det\left(H^\star_t\right)} \right) \\
    &\quad -\frac{1}{2}\sum_{t=1}^T {\widehat{\epsilon^\star}}'_t G_t^{-1} \widehat{\epsilon^\star}_t - \frac{1}{2} \sum_{i=1}^n {e_t^S}' S^{A'}_t \Psi^{-1} S^A_t e_t^S  - \frac{1}{2} \sum_{i=1}^n  {\ti{y}^\star_t}' \Sigma_t^{\theta Z}\ti{y}^\star_t , \\
    \{\Lambda\text{-terms}\} &= \frac{ns}{2} - \frac{1}{2}\sum_{i=1}^n\Tr\left(V_{\lambda_i}^{-1}\Sigma_{\lambda_i}\right) - \frac{1}{2}\sum_{i=1}^n{\mu_{\lambda_i}}'V_{\lambda_i}^{-1}\mu_{\lambda_i} -  \frac{1}{2}\sum_{i=1}^n\ln \left(\frac{\det\left(V_{\lambda_i}\right)}{\det\left(\Sigma_{\lambda_i}\right)}\right), \\
    \{\Phi\text{-terms}\} &= \frac{rs}{2} - \frac{1}{2}\sum_{j=1}^r\Tr\left(V_{\phi_j}^{-1}\Sigma_{\phi_j}\right) - \frac{1}{2}\sum_{j=1}^r{\mu_{\phi_j}}'V_{\phi_j}^{-1}\mu_{\phi_j} -  \frac{1}{2}\sum_{j=r}^n\ln \left(\frac{\det\left(V_{\phi_j}\right)}{\det\left(\Sigma_{\phi_j}\right)}\right), \\
    \{\Sigma_\epsilon\text{-terms}\} &= \sum_{i=1}^n\ln \Gamma\left(\frac{\nu_{\epsilon_i} + T_i}{2}\right) - \sum_{i=1}^n\ln \Gamma\left(\frac{\nu_{\epsilon_i}}{2}\right) - \sum_{i=1}^n\frac{\nu_{\epsilon_i} + T_i}{2}\ln\left( (\nu_{\epsilon_i} + T_i) \psi^2_{\epsilon_i}\right) + \sum_{i=1}^n\frac{\nu_{\epsilon_i}}{2}\ln \nu_{\epsilon_i} \tau^2_{\epsilon_i} \\
    &\qquad - \sum_{i=1}^n\frac{\nu_{\epsilon_i}\tau^2_{\epsilon_i}}{2\psi^2_{\epsilon_i}} + \sum_{i=1}^n \frac{\nu_{\epsilon_i} + T_i}{2},\\
    \{\Sigma_\ti{u}\text{-terms}\} &= \sum_{j=1}^r\ln \Gamma\left(\frac{\nu_{\ti{u}_j} + T}{2}\right) - \sum_{j=1}^r\ln \Gamma\left(\frac{\nu_{\ti{u}_j}}{2}\right) - \sum_{j=1}^r\frac{\nu_{\ti{u}_j}}{2}\ln\left( (\nu_{\ti{u}_j} + T) \psi^2_{\ti{u}_j}\right) + \sum_{j=1}^r\frac{\nu_{\ti{u}_j}}{2}\ln \nu_{\ti{u}_j} \tau^2_{\ti{u}_j} \\
    &\qquad - \sum_{j=1}^r\frac{\nu_{\ti{u}_j}\tau^2_{\ti{u}_j}}{2\psi^2_{\ti{u}_j}} + \sum_{j=1}^r \frac{\nu_{\ti{u}_j} + T}{2} -\frac{T}{2} \sum_{j=1}^r \ln \left(\frac{\nu_{\ti{u}_j} + T}{2}\right),  \\    
    \{Z\text{-terms}\} &= -\sum_{i=1}^n \sum_{k=1}^s b_{i,k} \ln \left(\frac{b_{i,k}}{\beta_{i,k}}\right) -\sum_{i=1}^n \sum_{k=1}^s (1-b_{i,k})\ln\left( \frac{1-b_{i,k}}{1-\beta_{i,k}}\right).
\end{align*} 

\end{appendix}
\end{document}